\documentclass[reprint,superscriptaddress,amsmath,amssymb,aps,floatfix]{revtex4-2}
\usepackage[english]{babel}
\usepackage[utf8]{inputenc}
\usepackage{hyperref}
\usepackage{siunitx} 
\usepackage{graphicx}
\usepackage{amssymb}
\graphicspath{{images/}{../images/}}
\usepackage[usenames, dvipsnames]{color}
\usepackage[table]{xcolor}

\usepackage{caption}
\usepackage{subcaption}

\begin{document}
\title{Laser Interactions with Gas Jets: EMP Emission and Nozzle Damage}
\author{P.~W.~Bradford} 
\email{philip.bradford@stfc.ac.uk}
\affiliation{Centre Lasers Intenses et Applications, University of Bordeaux-CNRS-CEA, 33405 Talence, Bordeaux, France}
\affiliation{Current affiliation: Central Laser Facility, Rutherford Appleton Laboratory, Didcot OX11 0QX, United Kingdom}
\author{V.~Ospina-Bohórquez}
\affiliation{Centre Lasers Intenses et Applications, University of Bordeaux-CNRS-CEA, 33405 Talence, Bordeaux, France}
\affiliation{CEA, DAM, DIF, F-91297 Arpajon, France}
\affiliation{Universit\'e Paris-Saclay, CEA, LMCE, 91680 Bruyères-le-Châtel, France}
\author{M. Ehret}
\affiliation{Centro de Láseres Pulsados, Building M5, Parque Científico, 37185 Villamayor, Salamanca, Spain}
\author{J.-L.~Henares}
\affiliation{Centro de Láseres Pulsados, Building M5, Parque Científico, 37185 Villamayor, Salamanca, Spain}
\author{P.~Puyuelo-Valdes}
\affiliation{Centro de Láseres Pulsados, Building M5, Parque Científico, 37185 Villamayor, Salamanca, Spain}
\author{T. Chodukowski}
\affiliation{Institute of Plasma Physics and Laser Microfusion, 23 Hery St., 00-908 Warsaw, Poland}
\author{T. Pisarczyk}
\affiliation{Institute of Plasma Physics and Laser Microfusion, 23 Hery St., 00-908 Warsaw, Poland}
\author{Z. Rusiniak}
\affiliation{Institute of Plasma Physics and Laser Microfusion, 23 Hery St., 00-908 Warsaw, Poland}
\author{C. Salgado-López}
\affiliation{Centro de Láseres Pulsados, Building M5, Parque Científico, 37185 Villamayor, Salamanca, Spain}
\author{C.~Vlachos}
\affiliation{Centre Lasers Intenses et Applications, University of Bordeaux-CNRS-CEA, 33405 Talence, Bordeaux, France}
\author{M.Scisci\`{o}}
\affiliation{ENEA - Nuclear Department, C.R. Frascati, Frascati, Italy}
\author{M. Salvadori}
\affiliation{ENEA - Nuclear Department, C.R. Frascati, Frascati, Italy}
\affiliation{Current affiliation: Consiglio Nazionale delle Ricerche, Istituto Nazionale di Ottica, CNR-INO, Pisa, 56124, Italy}
\author{C. Verona}
\affiliation{Dipartimento Ing. Industriale, Universita di Roma ``Tor Vergatà", Via del Politecnico 1, Roma 00133, Italy}
\author{G. S.~Hicks}
\affiliation{The John Adams Institute for Accelerator Science, Blackett Laboratory, Imperial College London, London, SW7 2BW, UK}
\author{O. C.~Ettlinger}
\affiliation{The John Adams Institute for Accelerator Science, Blackett Laboratory, Imperial College London, London, SW7 2BW, UK}
\author{Z. Najmudin}
\affiliation{The John Adams Institute for Accelerator Science, Blackett Laboratory, Imperial College London, London, SW7 2BW, UK}
\author{J.-R. Marqu\`es}
\affiliation{LULI, CNRS, \'{E}cole Polytechnique, CEA, Sorbonne Universit\'{e}, Institut Polytechnique de Paris, F-91128 Palaiseau Cedex, France}
\author{L.~Gremillet}
\affiliation{CEA, DAM, DIF, F-91297 Arpajon, France}
\affiliation{Universit\'e Paris-Saclay, CEA, LMCE, 91680 Bruyères-le-Châtel, France}
\author{J.~J.~Santos}
\affiliation{Centre Lasers Intenses et Applications, University of Bordeaux-CNRS-CEA, 33405 Talence, Bordeaux, France}
\author{F.~Consoli}
\affiliation{ENEA - Nuclear Department, C.R. Frascati, Frascati, Italy}
\author{V.~T.~Tikhonchuk}
\affiliation{Centre Lasers Intenses et Applications, University of Bordeaux-CNRS-CEA, 33405 Talence, Bordeaux, France}
\affiliation{Extreme Light Infrastructure ERIC, ELI Beamlines Facility, 25241 Dolní Břežany, Czech Republic}

\begin{abstract}
Understanding the physics of electromagnetic pulse emission and nozzle damage is critical for the long-term operation of laser experiments with gas targets, particularly at facilities looking to produce stable sources of radiation at high repetition rate. We present a theoretical model of plasma formation and electrostatic charging when high-power lasers are focused inside gases. The model can be used to estimate the amplitude of gigahertz electromagnetic pulses (EMPs) produced by the laser and the extent of damage to the gas jet nozzle. Looking at a range of laser and target properties relevant to existing high-power laser systems, we find that EMP fields of tens to hundreds of kV/m can be generated several metres from the gas jet. Model predictions are compared with measurements of EMP, plasma formation and nozzle damage from two experiments on the VEGA-3 laser and one experiment on the Vulcan Petawatt laser. 
\end{abstract}

\maketitle
\date{\today}

\section{Introduction}
High-power laser pulses focused into dense gases are used for the acceleration of charged particles and the generation of hard x-ray radiation, but these interactions also produce undesirable secondary effects, such as nozzle damage and the emission of electromagnetic pulses (EMPs). Laser-driven EMPs are produced in the radio-frequency domain and couple to motors, computers and other electronic equipment. On the VEGA-3 laser system \cite{Mendez_2019} at CLPU, EMPs have been responsible for valve malfunctions and gas leaks from jet nozzles; they are also known to enter diagnostics and oscilloscopes - ruining measurements of charged particle emission. 

Megahertz- and gigahertz-frequency EMPs are generated in a variety of high-power laser experiments when hot electrons are expelled from the target and oscillating currents are excited in the target mount and surrounding chamber \cite{Dubois2014}. Previous research \cite{Consoli_2020, Eder_2009, Brown_2013, Mead_2004, Poye_dynamic} has focused on laser interactions with solid targets, where the EMP amplitude is known to be the highest. The few measurements available for high-density gas jet targets \cite{Kugland_APL_2012, Micha_ECLIM_2019}, however, suggest that EMP emission is significant. If the EMP amplitude from gas jets scales with laser energy and intensity as for solid targets, these EMP fields will increase with a new generation of ultra-intense, high-repetition-rate laser systems \cite{Danson_2019}. 

\begin{figure}[!ht]
\centering
\includegraphics[width=0.48\textwidth]{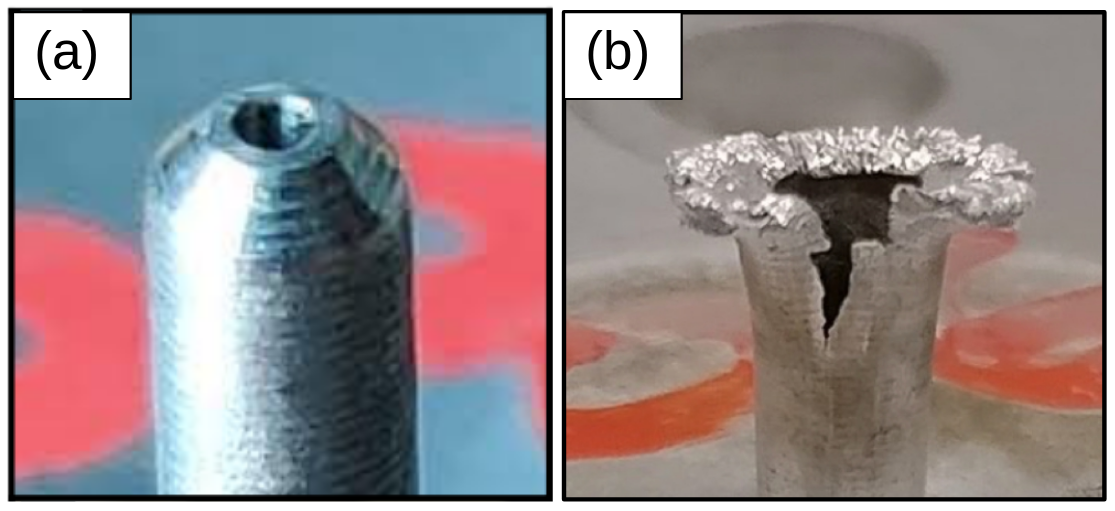}
\caption{Gas jet nozzle used during an experiment at Vulcan Target Area Petawatt (a) \textit{before} and (b) \textit{after} a full-power laser shot. Melting can lead to occlusion of the nozzle aperture or even total rupture of the material. Images reproduced from Hicks \textit{et al.} \cite{Hicks_2021} with permission.}
\label{fig_Hicks_nozzles}
\end{figure}

A second important concern of the laser-gas interaction is damage to the gas jet nozzle. Many of the most exciting applications of laser-gas research rely on high shot rates and a reproducible gas density profile to generate bright, high-fluence sources of energetic ions 
\cite{Lancaster_2004, Patel_2003, Nemoto_2001, Yogo_2011, Macchi_2013, Wei_2004}, electrons\cite{Wood_2018, Tsung_2004} or x-rays \cite{Barberio_2019, Ta_Phuoc_2012}. Repeated melting of the gas nozzle, therefore, represents a serious and expensive hindrance to this research. On energetic systems like the Vulcan Petawatt laser \cite{Danson_2004, Danson_2005} at the Rutherford Appleton Laboratory, a single shot on a dense gas target is sufficient to destroy the gas nozzle completely (see Fig. \ref{fig_Hicks_nozzles}). For systems operating at lower energy, lower gas density and shorter pulse duration, nozzle damage is more progressive but still leads to significant smoothing of the gas density profile \cite{Henares_2023, Ospina_2022,  Ospina_2024}, degradation of the laser interaction and reduced data reproducibility. Understanding the nozzle damage mechanism is therefore important for future laser-gas applications.

To date, no theoretical framework has been proposed for the emission of EMP from gas jets and this makes it difficult to estimate the severity of its impact on new laser systems. In Secs. \ref{sec_model} and \ref{sec_ionisation}, we present a model of laser-plasma expansion in a gas, where ionisation is mainly caused by ions streaming from the laser focus (anode) to the nozzle (cathode). In Sec. \ref{sec_spark_discharge}, however, we show that electrical breakdown induced by a strong plasma-nozzle potential can fully ionise the gas on shorter timescales of $10-20$~ps. EMP is then emitted similarly to solid-target interactions, where a discharge current propagates to ground (the chamber) along a cm-scale antenna (the nozzle or target support), emitting radiation at gigahertz frequencies. According to our model, the EMP amplitude for Vulcan Petawatt (Vulcan-TAP) and VEGA-3 interactions can reach tens to hundreds of $\mu$T or several tens of kV/m at a distance $\sim 1$~m from the target. The theoretical model is supported in Sec. \ref{sec_experiment} by the results from two experiments on the VEGA-3 laser, which show reasonable agreement with the discharge time, magnetic field spatial profile and accumulated target charge. 

In Sec. \ref{sec_nozzle_damage}, we consider how the nozzle may be damaged using the plasma expansion model. Two mechanisms are considered: (i) a kA-level discharge current and (ii) ion collisional heating. We show that damage to gas jet nozzles is more likely caused by the impact of plasma ions than by Ohmic heating of the nozzle surface. Model predictions for the EMP field strength and nozzle damage on different facilities is discussed in Sec. \ref{sec_discussion}. Finally, in Sec. \ref{sec_conclusion}, we present ideas for how the EMP emission model can be reliably benchmarked with simulations and dedicated experiments. The practical impact of this work is broad: allowing scientists to reduce damage to expensive gas jet nozzles and minimize the electrical disruption of equipment. It likewise represents a rich seam of more fundamental research, connecting the physics of laser-target charging, ionisation, ion acceleration, high-voltage breakdown and antenna emission. 

\section{Expansion of a Laser-Plasma in a Gas Jet} \label{sec_model}

We consider an expanding plasma created as a laser pulse propagates through a high-density gas. The laser pulse ionises the gas and creates a plasma channel. Plasma electrons are heated by laser radiation to relativistic energies, and some of them escape the channel and leave it positively charged. This charged plasma cylinder then expands into the surrounding gas or plasma, depending on the efficiency of the ionisation mechanisms discussed in Sec. \ref{sec_ionisation}. If the plasma expands into a gas, the gas is ionised until contact is made with the conducting tip of the gas nozzle. Once contact is made, and the plasma is connected directly to ground, a discharge is triggered and EMP radiation is emitted. Nozzle damage is determined variously by the amount of energy stored in the plasma, the strength of the discharge and the nozzle material.  

\begin{figure}[ht]
\centering
\includegraphics[width=0.48\textwidth]{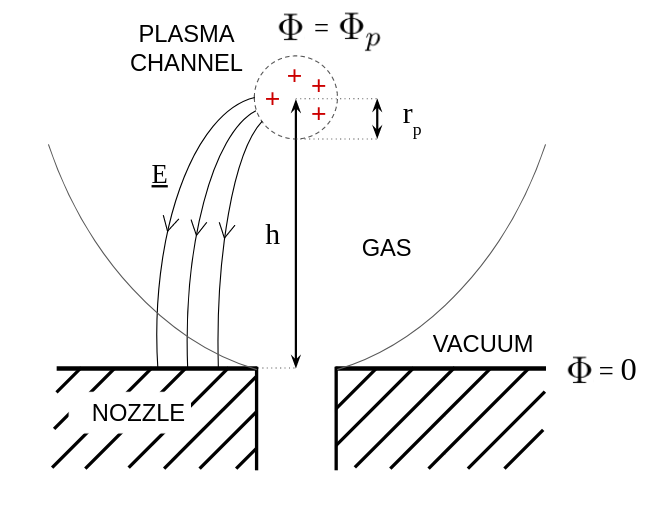} 
\caption{Schematic of the nozzle, gas and the cylindrical plasma channel formed by the laser. Here, $r_p$ is the radius of the plasma cylinder, $h$ is the separation between the channel and the nozzle, and $\Phi_p$ is the electric potential. The laser is directed into the plane of the page, along the axis of the plasma. Red crosses indicate a region of positive charge inside the laser-generated channel. Three curved arrows sketch the geometry of the electric field, \underline{E}. }\label{nozzle_schematic}
\end{figure}

Depending on the nozzle design and envisioned application, the laser pulse can be sent parallel to the surface of the nozzle tip at a height varying from a few tenths of a millimetre to a few millimetres. The width of the laser channel also depends on the laser focusing optics, laser power, and gas density profile along the laser axis. Figure \ref{nozzle_schematic} gives a schematic overview of the situation.

\subsection{Model of Electric Charging of the Plasma}
\label{plasma_params}

The model of EMP emission proposed in this paper consists of a phase of plasma charging, followed by a discharge and antenna emission process. First, the plasma is formed, and the charge is lost as hot electrons escape the potential barrier. This charge supplies the electrostatic energy that is later radiated as EMP. The mechanism of EMP emission, where the plasma is discharged through the nozzle and radiates as a dipole antenna, is similar to the solid-target model already examined in various publications \cite{Consoli_2020, Minenna_2020, Dubois2014, Poye_dynamic}.

Following the available observations and numerical simulations \cite{Debayle_2017, Vladisavlevici_2023, Krushelnick_1999, Willingdale_PRL_2006}, we assume that the plasma in the laser channel is fully ionised. The absorbed laser energy $\eta_L E_L$ is transferred to electrons with an energy distribution approximated by a two-temperature relativistic distribution:
\begin{equation}\label{eq1a}
    f_e(\gamma)=n_e f_{\rm MJ}(\gamma,\Theta_e) +n_h f_{\rm MJ}(\gamma,\Theta_h),  
\end{equation}
where $f_{\rm MJ}(\gamma,\Theta)$ is the Maxwell-J\"uttner function\cite{Passoni_2010}
$$ f_e(\gamma,\Theta)=(\gamma/\Theta)\, K_2(1/\Theta) \sqrt{\gamma^2-1}\exp(-\gamma/\Theta) $$
depending on the dimensionless temperature $\Theta= T/(m_ec^2)$, the modified Bessel function $K$ of the second kind and the electron relativistic factor $\gamma=1+ \varepsilon/(m_ec^2)$.
A fraction $\eta_{\rm las\to h}$ of the absorbed laser energy is transferred to a population of hot electrons with density $n_h$ and temperature $T_h$, while the remaining energy goes to electrons in the plasma bulk, with density $n_e$ and temperature $T_e$. Here, we use subscripts ``$h$" and ``$e$" to refer to the hot and thermal electron populations respectively.

The bulk electron density is defined by the gas density ($n_e = Z n_{\rm at}$) and the bulk temperature is defined by the plasma volume and deposited energy via: 
\begin{equation}\label{eq1b}
C_e(T_e) n_e T_e =(1-\eta_{\rm las\to h})\eta_L E_L/V_p,
\end{equation}
where $C_e (T_e)= 3-[1- K_1(1/\Theta_e)/ K_2(1/\Theta_e)]/\Theta_e$ is the heat capacitance of a relativistic electron gas. 

The hot electron density can then be defined knowing the fraction of energy transferred to hot electrons $\eta_{\rm las\to h}$: 
$$ C_e(T_h) n_h T_h =\eta_{\rm las\to h}\eta_L E_L/V_p. $$

The electric potential of the plasma channel is determined by how many hot electrons can overcome the plasma potential barrier. Though electrons from the plasma bulk will also contribute to the ejected charge, their contribution is less than $10\%$ because a significant fraction of the available laser energy is transferred to hot electrons. The electric potential $\Phi_p$ at the surface of the plasma cylinder with respect to the grounded nozzle can be estimated as
\begin{equation}\label{eq1}
\Phi_p = Q/C,
\end{equation}
where $Q$ is the charge of the plasma and $C= 2\pi l_p\epsilon_0/\ln(h/r_p)$ is the capacitance. The capacitance follows from the potential of a charged cylinder, radius $r_p$, length $l_p$, situated a distance $h$ from an infinite conducting plane. Only electrons with energies $\varepsilon_e>e\Phi_p$ can escape plasma to the ground. Assuming, as mentioned above, that hot electrons have an exponential distribution in energy with a temperature $T_h$, the fraction of escaped hot electrons is \cite{Chen_1984}: 
\begin{equation}\label{eq2}
\frac{\delta n_h}{n_h}= \int_{1+\psi}^\infty d\gamma\, f_{\rm MJ}(\gamma, \Theta_h),
\end{equation}
where $\psi=e\Phi_p/(m_ec^2)$ is the dimensionless potential. The plasma charge is then $Q= e \delta n_h  V_p$ and, substituting this expression into Eq. \eqref{eq1}, we have the following equation for the dimensionless potential $\psi$:
\begin{equation}\label{eq3}
\psi = \frac{r_p^2 \omega_{ph}^2}{2c^2} \,\frac{\delta n_h}{n_h}\, \ln\frac{h}{r_p},
\end{equation}
where $\omega_{ph}=(e^2 n_h/m_e\epsilon_0)^{1/2}$ is the hot electron plasma frequency.

The total energy of escaped electrons is split between the electrostatic energy of the charged plasma $\mathcal{E}_{\rm es}=Q^2/ 2C$ and the kinetic energy of the escaped electrons:
\begin{equation}
     \mathcal{E}_{\rm esc}= V_p n_h m_e c^2 \int_{1+\psi}^\infty d\gamma\,(\gamma-1) f_{\rm MJ}(\gamma, \Theta_h). \label{eq4}
\end{equation}
 
\subsection{Ion Acceleration in the Expanding Plasma}\label{ion-accel}

The number of ions in the laser plasma and their energy are important factors for determining the extent of damage to the gas nozzle, as well as for placing a lower limit on the speed of the plasma discharge. When the expanding thermal ions reach the nozzle surface they deposit their kinetic energy and can cause the nozzle to melt. Contact between the plasma and the nozzle also establishes an electrical path to ground, triggering a plasma discharge and the emission of EMP radiation.

The ion spectrum can be separated into two broad populations \cite{Puyuelo_Valdes_2019, Ospina_2024}: (i) thermal ions that expand as part of the laser plasma and (ii) a less numerous population of fast ions accelerated in the charge separation field produced by the escaped hot electrons \cite{Sentoku_2000,Debayle_2017}. We restrict ourselves to modelling the thermal ions, since they carry most of the energy.

Following a short phase of gas ionisation, electron heating and plasma charging, the plasma expands and cools down. After the end of the laser pulse, there is no more energy supply and the plasma expands adiabatically under the electron thermal pressure. The plasma pressure $p_e = n_e T_e$ is much higher than the ambient gas pressure, so the plasma expands freely with electrons transferring their energy to ions. The plasma charge is conserved during the expansion phase and the potential decreases logarithmically as the plasma radius increases (see Eq. \eqref{eq3}). The plasma expansion can be described in some special cases by a self-similar rarefaction wave model \cite{Mora_2003, Murakami_2006}. More detailed analysis is performed for a spherical plasma expansion in Ref. \cite{Popov_2010}. In practice, the ion energy distribution depends on the density profile and on the ratio of the plasma radius to the Debye length. For our purposes, it is sufficient to estimate the average ion kinetic energy $\varepsilon_i$ by equating the ion and electron energy densities, $n_i \varepsilon_i \simeq n_e \varepsilon_e \approx \frac{3}{2} n_e T_e$, where $n_{i}$ is the ion density in the plasma channel and $\varepsilon_e $ is the average electron kinetic energy. The total number of ions in the plasma is $N_i = n_i V_p$.

\subsection{Model estimates for Vulcan-TAP conditions} \label{sec_Vulcan_params}

\begin{table}
\begin{center} \begin{tabular}{|c || c|c|} 
 \hline
 \multicolumn{3}{|c|}{Laser Parameters} \\ \hline
 \rowcolor{lightgray} Parameter & Vulcan-TAP & VEGA-3 \\ [0.5ex]  \hline
 $I_L$ [W~cm$^{-2}$] & $2.9\times10^{20}$ & $3.6\times10^{19}$ \\
 $E_L$ [J] & 80 & 8 \\ 
 $\tau_L$ [fs] & 600 & 75 \\
 $w_L$ [$\mu$m] & 3.9 & 9.8 \\
 $\lambda_L$ [$\mu$m] & 1.05 & 0.8 \\
 $f_{\#}$ & 3 & 10 \\
 $a_0$ & 15 & 4.1 \\ \hline
\end{tabular}
\end{center}
\caption{Representative input laser parameters - measured or inferred - as described in the text and Refs. \cite{Hicks_2021, Ospina_2024}.}\label{tab1a}
\end{table}

\begin{table}
\begin{center} \begin{tabular}{|c || c|c|} 
 \hline
 \multicolumn{3}{|c|}{Gas and Plasma Parameters (Measured)} \\ \hline
 \rowcolor{lightgray} Parameter & Vulcan-TAP & VEGA-3 \\ [0.5ex] 
 \hline
 Gas & H$_2$ & He \\ 
 $n_e$ [cm$^{-3}$] & $8.1\times10^{20}$ & $2.4\times10^{20}$ \\ 
 $p_{\rm gas}$ [bar] & 15 & 4.5 \\
 $h$ [$\mu$m] & 500 & 500 \\
 \hline
 \multicolumn{3}{|c|}{Gas and Plasma Parameters (Assumed)} \\
 \hline
 $r_p$ [$\mu$m] & 20 & 20 \\ 
 $l_p$ [$\mu$m] & 500 & 500 \\
 $\eta_L $ & 1.0 & 0.9 \\
 $\eta_{\rm las\to h}$ & 0.4 & 0.1 \\
 \hline
\end{tabular}
\end{center}
\caption{Representative gas and plasma parameters (measured and assumed). The plasma radius $r_p$ and the laser energy fraction converted to hot electrons $\eta_{\rm las\to h}$ are estimated from dedicated PIC simulations\cite{Hicks_2021,Ospina_2024}.}\label{tab1b}
\end{table}

\begin{table}
\begin{center} \begin{tabular}{|c || c|c|}  \hline
 \multicolumn{3}{|c|}{Plasma Properties} \\
 \hline
 \rowcolor{lightgray} Parameter & Vulcan-TAP & VEGA-3 \\ [0.5ex] 
 \hline
 $V_p$ [cm$^3$] & $6\times10^{-7}$ & $6\times10^{-7}$ \\
 $T_e$ [keV] & 280 & 140 \\ 
 $T_h$ [MeV] & 5.0 & 1.1 \\ 
 $\Phi_p$ [MV] & 25 & 4.4 \\
 $Q$ [nC] & 230 & 42 \\ [0.5ex]
 \hline
\end{tabular}
\end{center}
\caption{Plasma properties estimated using the model presented in Sec. \ref{plasma_params}.}\label{tab1c}
\end{table}

The plasma charging model described in the previous section takes several laser and gas parameters as inputs which must either be experimentally determined or estimated by other means. Input parameters relevant to the experiments considered in this paper are given in Tabs. \ref{tab1a} and \ref{tab1b}, along with the outputs from our model given in Tab. \ref{tab1c}. We use Vulcan-TAP parameters in Secs. \ref{sec_ionisation} and \ref{sec_nozzle_damage} to explore likely mechanisms of gas ionization and nozzle damage. The VEGA-3 parameters are used to benchmark our model against the experiments described in Sec. \ref{sec_experiment}. 

We assume the laser energy is separated as follows: the total laser energy is multiplied by some fraction to reflect the amount of energy contained within the laser focus, which is called $E_L$; the energy in the focus is then multiplied by a fraction $\eta_L$, which reflects its absorption in the gas and a further fraction $\eta_{\rm las\to h}$, which is the fraction of energy converted to hot electrons. The laser energy not converted into hot electrons goes into the thermal electrons and is eventually converted into the kinetic energy of the plasma ions as the plasma expands.

Four of the input parameters for our model are ``assumed", which means they have been estimated based on simulations or previous experimental data. The laser absorption $\eta_L$ in the gas is inferred from dedicated PIC simulations \cite{Ospina_2024, Hicks_2021}, which indicate that the laser is entirely depleted after it has passed through the dense gas. The conversion efficiency of laser energy to hot electrons, $\eta_{\rm las\to h}$, is a function of many variables, including the gas density and laser intensity. It is not measured in any of the experiments described here and must therefore be estimated to the nearest order of magnitude. For near-critical density plasmas the available data is relatively scarce, though there is evidence that the hot electron conversion efficiency ranges from a few percent to more than ten percent under certain conditions \cite{Brunetti_2022,Feng_2023}. Here, we take an upper estimate of $\eta_{\rm las\to h}=0.4$ to illustrate a scenario of strong charging and EMP. The plasma channel radius and length are estimated from PIC simulations \cite{Ospina_2024, Hicks_2021, Willingdale_PRL_2006, Bonvalet_2021} and experimental interferograms \cite{Ospina_2024, Maitrallain_2024}, which suggest that channels several hundred microns in length and tens of microns in radius can be formed on the Vulcan-TAP and VEGA-3 laser systems a few ps after the arrival of the laser pulse. 


Consider conditions on the Vulcan-Petawatt laser, with pulse energy $E_{\rm L,tot}=200$~J, duration $\tau_L=600$~fs and wavelength $\lambda_L=1.053\,\mu$m \cite{Hicks_2021}. The laser beam is focused using $f_\#=3$ optics to a spot with full width at half maximum (FWHM) $w_L=3.9\,\mu$m. Only 40\% of the total laser energy is contained within the FWHM, for a maximum on-target intensity of $I_L\simeq 2.9\times 10^{20}$~W~cm$^{-2}$. The spatial extent of the laser channel in the gas depends on the wavelength of the laser and the gas density. For typical gas pressures $p_{\rm gas}$ ranging from 0.03 to 30~bar, the atomic/molecular densities in the laser focal region are in the range  $n_{\rm gas}\sim 10^{18}-10^{21}$~cm$^{-3}$. The maximum electron density in a fully ionised plasma is smaller than the critical density, so the laser can propagate through the gas. 

The Rayleigh length of a diffraction-limited Gaussian beam in our example is $Z_R = \pi w_L^2 / \lambda_L \simeq 44\,\mu$m. Since the gas jet width is usually larger than $Z_R$ and the beam is susceptible to relativistic self-guiding, the laser produces a channel significantly longer \cite{Maitrallain_2024, Sylla_2013} than $Z_R$. It can be viewed as a $l_p\simeq 0.5$~mm-long plasma cylinder created at a typical height of $h\simeq 0.5$~mm above the nozzle. We take the initial radius of the plasma channel to be a factor of a few times larger than the laser focal radius, $r_p=20\,\mu$m, to reflect the average size of the channel as the laser deposits energy over the full length of its path through the gas. We assume that laser absorption in the gas jet is $\eta_L\sim 100$\%, with 40\% transferred to hot electrons.

For the parameters of the Vulcan experiment given in Tabs. \ref{tab1a} - \ref{tab1c}, a plasma channel of radius $r_p=20\,\mu$m and length $l_p=0.5$~mm has volume $V_p=\pi r_p^2 l_p \simeq 6\times 10^{-7}$~cm$^{3}$. A hydrogen gas pressure of 15~bar corresponds to an atomic gas density $n_{\rm at} \simeq 8.1 \times 10^{20}$~cm$^{-3}$ at a distance $h\simeq 0.5$~mm above the nozzle, so we have $N_e\simeq 5\times 10^{14}$ electrons. Equation \eqref{eq1b} yields an effective electron temperature $T_e= 250$~keV for the total bulk energy of 43~J. 

A hot electron temperature of $T_h=5.0$~MeV is prescribed following the ponderomotive scaling \cite{Wilks_1992}, $(\sqrt{1  + a_0^2} - 1)m_ec^2$. If we further assume that $\eta_{\rm las\to h} = 0.4$ on Vulcan, then Eq. \ref{eq2} gives a hot electron fraction $n_h/n_e \simeq 2.5$\% with total energy 29~J. The hot electron density follows from the hot electron temperature and the heat capacity of a relativistic electron gas, as shown in the previous section.


The solution of Eq. \eqref{eq3} for typical parameters $h/r_p= 25$ and $r_p\omega_{ph}/c= 18$ gives $\Phi_p =24.8$~MV. Knowing the plasma potential and the plasma capacitance $C\simeq 0.0086$~pF, we find the charge ejected from the plasma column, $Q =C \Phi_p \simeq 234$~nC. This corresponds to a fraction of escaped electrons $\delta n_h/n_e \approx 2.8\times10^{-3}$ and is consistent with the experiment \cite{Simpson_2021}. 

Since the time required for a hot electron to cross the channel is relatively short $\sim 2r_p/c= 0.13$~ps, the time of plasma charging is determined by the propagation time of the laser pulse through the gas, $\sim 3$~ps.

Following Eq. \eqref{eq4}, the total energy of escaped electrons $\mathcal{E}_{\rm esc}$ is divided between the electrostatic energy of the charged plasma $\mathcal{E}_{\rm es}=Q^2/ 2C \simeq 3.2$~J and the kinetic energy of the escaped electrons $\simeq 4.2$~J. 

As discussed in the previous section, the average ion kinetic energy, $\varepsilon_i \simeq 0.4$~MeV, is 1.5 times the bulk electron temperature for a hydrogen plasma. This corresponds to an ion velocity of $v_i\simeq 9\,\mu$m/ps, and an ion expansion time to the nozzle $t_{\rm exp}\simeq 60$~ps. The total energy carried by these ions is of order the energy carried by the expanding thermal electrons - that is, about $(1 - \eta_{\rm las\to h})\eta_L E_L \sim 40$~J for the Vulcan experiment. 

\begin{figure}[!ht]
\centering
\includegraphics[width=0.35\textwidth]{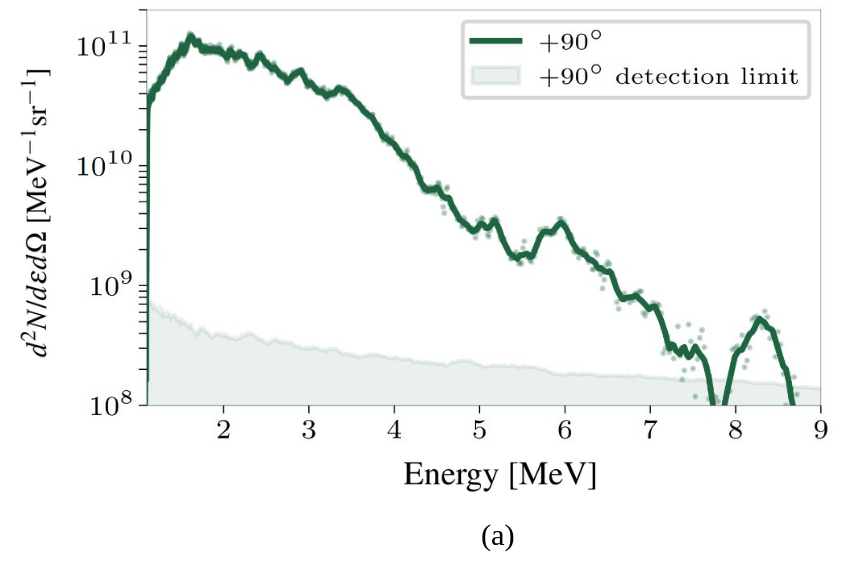} \\
\includegraphics[width=0.35\textwidth]{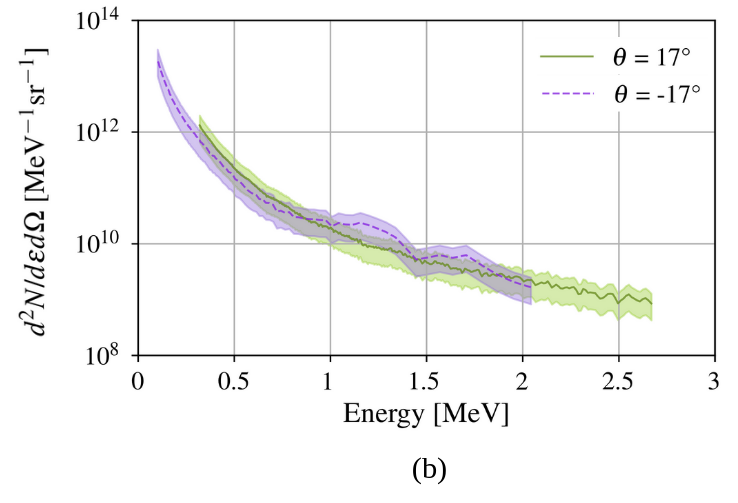} \\
\includegraphics[width=0.35\textwidth]{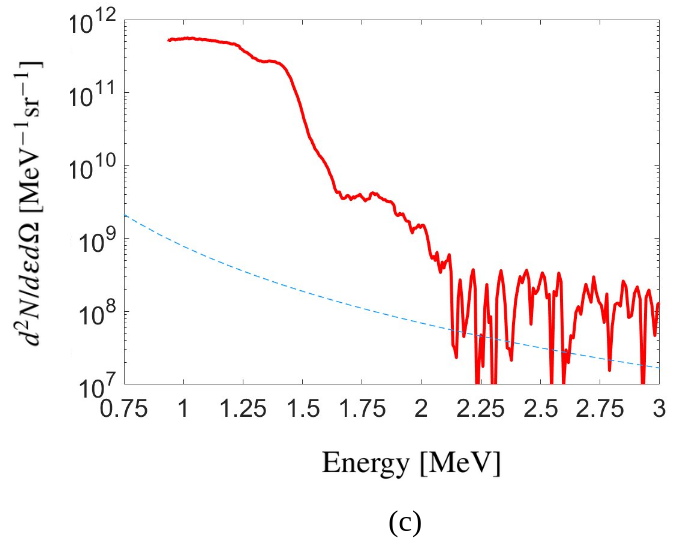} 
\caption{Ion spectra collected in gas jet experiments: (a) Proton spectrum from an experiment on Vulcan-TAP, measured at $90^\circ$ to the laser axis using a Thomson parabola spectrometer with BAS-TR image plate \cite{Hicks_2021}. The shaded region is the $3\sigma$ detection limit. (b) $\alpha$-particle spectra from an experiment at VEGA-3 \cite{Ospina_2024}, measured at $\pm17^{\circ}$ from the laser axis using Diamond time-of-flight detectors. The two spectra were recorded on different shots. See also Fig. \ref{fig_VEGA3_expts}a. (c) Proton spectrum from a separate experiment at VEGA-3 \cite{Henares_2023}, measured using a Thomson parabola spectrometer at $90^\circ$ to the laser axis. The spectrometer dynamic range limits reliable measurements to energies $\gtrsim1.3$~MeV. The blue dashed line represents the background noise level. See also Fig. \ref{fig_VEGA3_expts}b.}
\label{fig_ion_spectra}
\end{figure}

The model estimate of $5\times10^{14}$ ions with a combined kinetic energy of several tens of joules can be compared with measured ion spectra and simulations. Fig. \ref{fig_ion_spectra}a shows an ion spectrum from Ref. \cite{Hicks_2021} measured at $90^\circ$ to the laser axis. The peak ion yield of $\sim 10^{11}$~MeV$^{-1}$sr$^{-1}$ occurs at $\sim 2$~MeV. The proton signal drops off at lower energies because the configuration of B-/E-fields inside the spectrometer limits its dynamic range to about two decades. Other ion spectral measurements \cite{Marques_2021, Marques_2023, Puyuelo_Valdes_2019, Ospina_2024, Wei_2004} and simulations \cite{Bonvalet_2021, Hicks_2021} suggest that the sub-MeV thermal ion population should be orders of magnitude more numerous than the multi-MeV plasma-accelerated ions. Ion spectra measured at the PHELIX laser using spectrometers specifically designed to measure sub-MeV ions \cite{Marques_2021} show sub-MeV ion yields of $\sim 10^{14}$ particles.

In addition to ions accelerated by bulk electron pressure in the expanding plasma, some ions are accelerated to energies of a few MeV in the electrostatic field created by the escaped electrons. Hicks \textit{et al.} \cite{Hicks_2021} report on the number of such fast protons accelerated in the radial direction as 30~nC/sr. Assuming an emission solid angle of $1-2$~sr, the total charge of ejected protons is about 60~nC and their overall energy about 0.2~J. These values are consistent with 230~nC charge of ejected electrons and a corresponding energy of $\sim 4$~J, as estimated above. 

\section{Gas Ionisation by Laser-Plasma Products} \label{sec_ionisation}

The level of EMP emission depends on the intensity of the discharge current and, consequently, the resistance of the plasma that connects the laser channel to the nozzle. There are four processes of gas ionisation: (i) photoionisation from UV and x-rays produced by the bremsstrahlung of hot electrons in the plasma column, (ii) collisional and field ionisation from fast electrons accelerated in the laser channel, (iii) collisional ionisation from fast ions emanating from the laser channel and (iv) ionisation by electron avalanche caused by a high plasma-nozzle electric potential. 

\subsection{X-ray Photoionisation of the Gas} \label{sec_photoionisation}
The bremsstrahlung emission of electrons in the laser-created plasma covers a broad range of photon energies up to the electron kinetic energy. Assuming the electron energy distribution is characterized by an effective temperature $T_e$, the power per unit volume of bremsstrahlung emission from a hydrogen plasma can be written as \cite{NRL_2019}:
$$ P_{\text{Brem}} = 1.69\times10^{-32} n_e^2 T_e^{1/2}\,{\rm W~cm}^{-3},$$
where the electron density is in cm$^{-3}$ and the electron temperature is in eV. The radiative contribution of bulk electrons dominates that of hot electrons because $n_e T_e \gtrsim n_h T_h$ under the present conditions. For an electron density of $n_e=8\times10^{20}$~cm$^{-3}$ in fully ionised plasma and a temperature $T_e\simeq 250$~keV, we have $P_{\text{Brem}} = 6\times 10^{12}$~W~cm$^{-3}$. Accounting for the plasma volume of $V_p \sim 6 \times 10^{-7}$~cm$^3$ and lifetime $t_{\rm exp}\sim 60$~ps (see Sec. \ref{sec_Vulcan_params}), the total radiated energy is about $\mathcal{E}_{\rm Brem}=P_{\rm Brem} V_p t_{\rm exp}\simeq 0.2$~mJ. The spectrum of emission is flat for photons with energy $\varepsilon_{x} \lesssim T_e$ and decreases exponentially with a temperature $T_{\rm rad} \simeq T_e$ for higher energies.   

The mean free path of photons with energies above 3~keV in hydrogen is approximately constant at $\lambda_x \rho_{\rm gas} \sim 3$~g~cm$^{-2}$\,\cite{Yao_2006}. For a gas jet density of $\rho_{\rm gas} \sim1.3$~mg cm$^{-3}$, the absorption length is 20~m and the probability of gas ionisation over a millimetre distance is negligible. The mean free path of lower energy photons strongly decreases as the energy decreases \cite{Yao_2006}. It can be approximated as $\lambda_x \rho_{\rm gas} \sim 10^{-10} \varepsilon_x^3$~g~cm$^{-2}$ for hydrogen, where $\varepsilon_x$ is the photon energy in eV. The mean free path of photons with the energy of $15-20$~eV - comparable to the hydrogen ionisation potential $U_{\rm ion}=13.6$~eV - is therefore just a few microns. 

The photons with energy about $\varepsilon_h\sim 100-200$~eV produce the most efficient photoionisation because their mean free path is comparable to the plasma-nozzle distance. The fraction of bremsstrahlung energy carried by these photons is $\varepsilon_h/T_e \sim 10^{-3}$ - that is, about $0.2\,\mu$J. Knowing the energy required to ionise a hydrogen atom and the volume of gas between the plasma and the nozzle, $V_{\rm gas}\sim 4\times 10^{-4}$~cm$^3$,  the gas ionisation level can be estimated as $\mathcal{E}_{\rm Brem} \varepsilon_h/T_e U_{\rm ion} V_{\rm gas} \sim 1.7\times 10^{14}$~cm$^{-3}$. This corresponds to an ionised fraction of approximately $3 \times 10^{-7}$. The photoionisation time of $\sim 60$~ps is defined by the time of bremsstrahlung emission, which is the lifetime of the hot plasma channel.

\subsection{Ionisation by Hot Electrons}

Hot electrons also contribute to the ionisation process. These electrons are those escaping the plasma potential barrier and their characteristics are estimated in Sec. \ref{sec_model}. The stopping power of electrons in hydrogen gas for this range of energies of a few MeV is $\sim 4$~MeV~cm$^2$/g \,\cite{ESTAR}, so a hot electron loses $\sim 260$~eV over the distance $h\simeq 0.5$~mm. The energy of secondary electrons is comparable to the ionisation potential of the gas \cite{Chepel_2013}, which means that one primary electron creates $\sim 20$ secondary electrons in the gas volume. Using the number of escaped electrons estimated in Sec. \ref{plasma_params}, $N_{\rm esc}\simeq 1.4\times 10^{12}$, the density of secondary electrons produced by collisional ionisation is $\sim7.5\times 10^{16}$~cm$^{-3}$. This corresponds to an ionised fraction of $\sim 10^{-4}$, which is produced in a short time of hot electron emission of a few ps.

Alternatively, hot electrons can produce field ionisation of the gas if the electron beam density is sufficiently high. For a beam of 3~MeV electrons passing through a neutral gas, field ionisation dominates over collisional ionisation for beam densities exceeding $\sim 10^{17}$~cm$^{-3}\,$ \cite{Tikhonchuk_2002}. Considering $10^{12}$ hot electrons contained within the initial plasma volume, the maximum beam density is $\sim 2.5\times 10^{17}$~cm$^{-3}$. So field ionisation has approximately the same impact as collisional ionisation.   

\subsection{Ionisation by Plasma Ions} \label{sec_ion_ion}

The stopping power of a representative 0.4~MeV proton in hydrogen is 1200~MeV cm$^2$/g \,\cite{PSTAR}. The energy needed for the creation of one secondary electron by a fast ion is of order the ionisation potential, $U_{\rm ion}$, and it varies weakly with the ion energy \cite{Giesen_2014, Baek_1994}. A 0.4~MeV proton passing through hydrogen gas, therefore, creates $\sim 6000$ secondary electrons over the 0.5~mm distance between the plasma and the nozzle, losing about 0.1~MeV of its kinetic energy. In a gas volume of $\sim 4\times 10^{-4}$~cm$^{-3}$, this corresponds to complete ionisation of the ambient gas. This is, however, a relatively long process that takes $\sim 60$~ps.

The contribution of fast ions to gas ionisation is smaller by more than three orders of magnitude, but their velocity is about three times higher. An ionisation level of $\sim 3\times 10^{-3}$ is therefore produced prior to plasma expansion, during the time of flight of the fast ions, which is about 20~ps.

\subsection{Gas Ionisation by Electrical Breakdown} \label{sec_spark_discharge}

When hot electrons are ejected from the laser plasma, it becomes positively charged with respect to the nozzle. The electric field between the plasma and nozzle may then break down the gas and propagate a discharge. The discharge regime is determined by the product of the gas pressure and electrode separation $p_{\rm gas}h$. Under Vulcan-TAP conditions, with $p_{\rm gas}h\gtrsim 1$~bar~cm, avalanche ionisation will occur if the potential is above the breakdown value of $20-30$~kV \,\cite{Raizer_1991}. This is much smaller than the plasma potential of $\sim 25$~MV estimated in Sec. \ref{sec_model}. 

In the avalanche process, the electron density increases exponentially in time from the seed level $n_{e0}/n_{\rm at}\sim 10^{-4}-10^{-3}$ created by fast electrons or ions, $n_e(t)=n_{e0}\, \exp(\nu_{\rm ion}t)$, where $\nu_{\rm ion}$ is the characteristic ionization rate. According to Refs. \cite{Raizer_1991, Solovyev_1999}, the ionisation rate is $\nu_{\rm ion}\sim 10^{12}$~s$^{-1}$ for Vulcan conditions and the gas can be fully ionised in a few picoseconds. However, the plasma potential decreases with time for two reasons: first, it decreases logarithmically with time because of plasma expansion according to Eq. \eqref{eq3} and second, because the energy deposited during the ionisation process is extracted from the electrostatic energy. The decrease in electrostatic energy is related to the ionisation loss, $\Delta \mathcal{E} \simeq U_{\rm ion} n_e V_{\rm gas}$, and the discharge stops when all the available electrostatic energy $\mathcal{E}_{es}$ has been exhausted. This relation defines the maximum ionisation that can be achieved by avalanche breakdown, $n_{e,\max}\sim 4\times 10^{21}$~cm$^{-3}$, which is larger than the atomic gas density.

\subsection{Conclusions on Ionization Processes and Plasma Resistance} \label{sec_resistance}

In conclusion, photoionisation of the gas jet is low and can be neglected. Instead, ionisation by charged particles is produced in three steps: first, at the level of $10^{-4}$ by fast electrons in a few ps, then at the level of $10^{-3}$ by fast ions within 20~ps, then finally by the expanding plasma on a timescale of 60~ps. Free electrons produced during the first two steps provide the seed for a discharge by electrical breakdown. Avalanche breakdown is initiated from the seed level of ionisation produced by fast electrons or fast ions on a $10-20$~ps timescale. It can fully ionise the gas under the conditions of the Vulcan experiment before the expanding plasma reaches the nozzle. 

The degree of gas ionisation determines the plasma resistance and the discharge current that can be supported. We estimate the plasma resistance as $R = \eta h/ A$, where $\eta$ is the plasma resistivity and $A\approx 2 r_p l_p$ is the cross-sectional area of the ionised gas between the nozzle and the laser channel. The plasma resistivity in a fully ionised plasma is dominated by electron-ion collisions and can be estimated from the Drude formula:
\begin{equation} \label{Drude} 
\eta = m_e \nu_{ei}/e^2 n_e,
\end{equation}
where $\nu_{ei}$ is the electron-ion collision frequency. Assuming the electron temperature to be of the order of the ionisation potential, we find $\nu_{ei} \sim 10^{14}$~s$^{-1}$ and the plasma resistivity is  $\eta \simeq 5\times 10^{-6}\,\Omega$~m. This corresponds to a very small resistance of order $0.1\,\Omega$. A path to ground is established when the gas is fully ionised; the potential drop across the nozzle is of comparable magnitude to the plasma potential and a discharge current is produced. The accumulated plasma charge and discharge time puts a limit on the maximum discharge current of order $\sim10$~kA, as discussed in the next section.

\section{EMP emission} \label{sec_EMP_energy}

Here, we estimate the characteristics of EMP emission. The theoretical picture is as follows: the plasma supplies a current limited by the nozzle resistance and the available charge, and radiation is emitted from an antenna made out of the conducting parts of the nozzle. Considering the nozzle as a quarter-wavelength dipole, the emission frequency depends on the nozzle length $h_d$ as $\nu_{\rm emp}=c/4h_d$. For $h_d=3$~cm, the emission frequency is 2.5~GHz. 

The EMP energy is limited by the available electrostatic energy in the plasma, which is of order a few joules. Modelling the plasma-nozzle system as a dipole, the total emitted energy, according to the textbook by Jackson \cite{Jackson_2003}, depends on the charge $Q$ and the emission frequency: 
\begin{equation}\label{eq6}
 \mathcal{E}_{EMP} \approx 0.1 Z_0 Q^2 \nu_{\rm emp}
\end{equation}
where $Z_0=377$~$\Omega$ is the vacuum impedance. For 230~nC accumulated charge and an emission frequency of 2.5~GHz, the relation yields $\sim 5$~mJ of EMP energy and an average current at the antenna frequency \cite{Minenna_2020} of $J_{\text{emp}} = Q \nu_{\rm emp} \sim 580$~A. EMP emission is delayed with respect to the laser: first, because it takes time to ionise the gas and second because the current needs time to propagate along the antenna. 

Knowing the amplitude of the discharge current, we can also estimate the amplitude of the emitted magnetic field at a distance $r$ as $B\simeq  \mu_0 J_{\text{emp}} / 2 \pi r$, where $\mu_0$ is the permeability of free space. This is equivalent to the maximum magnetic field (measured over all possible emission angles) in the far-field of a dipole antenna \cite{Minenna_2020}. For a nozzle of height $h_d=3$~cm, the peak magnetic field amplitude is $\sim 120$~$\mu$T at a distance of $r=1$~m.
The gas discharge is thus expected to produce an electromagnetic pulse with electric field of amplitude $E=cB \approx 36$~kV/m (in the plane wave approximation). This value is higher than the electromagnetic susceptibility threshold for electronics \cite{Bardon_2020}.

\section{Nozzle Damage Mechanism} \label{sec_nozzle_damage}

We identify two mechanisms \footnote{UV, soft x-rays and thermal electrons can also couple to the nozzle and produce heating. Hard x-rays will deposit their energy volumetrically and are less important. The energy in UV and x-rays is relatively small, however (see Sec. \ref{sec_photoionisation}), while the thermal electrons travel with the ions and have a lower stopping power\cite{ESTAR}, producing heating in conjunction with the ions.} that could lead to nozzle damage in high-power laser experiments with gas jets: Ohmic heating from a discharge current or bombardment by plasma ions. To evaluate which process is more likely to induce melting, we calculate the energy required to melt a given mass of nozzle material. 

We consider copper as a representative nozzle material, with a heat capacity $C_V = 0.39$~J/gK, a melt temperature of 1360~K and a latent heat of melting 0.86~kJ/g. To heat 1~g of Cu to the melting temperature - and to melt it - one needs a total energy of 1.3~kJ. Similar estimates apply to aluminium and iron. The total available electrostatic energy in the plasma of 3~J can melt a maximum of 2~mg Cu. By contrast, the total kinetic energy of the expanding plasma ions is more than 10 times larger. The plasma can therefore cause more damage to the nozzle.

\subsection{Energy Released through Ohmic Heating}

The energy released by Ohmic heating is $J^2Rt_p$, where $J=Q/t_p$ is the discharge current. The resistance $R$ is calculated from the metal resistivity $\eta_d$ and the current-carrying surface $A_d$ and nozzle length $h_d$. Since the discharge is short, the current is confined within a skin depth, $\delta=\sqrt{\eta_d/(\nu_{\rm emp}\mu_0)}$. For Cu resistivity $\eta_d =1.7 \times 10^{-7}\,\Omega$~m and the frequency $\nu_{\rm emp} \approx 2.5$~GHz, the skin depth is $\delta\sim 10~\mu$m. Assuming the current flows along $h_d=3$~cm nozzle with a radius of $r_d=5$~mm, the current-carrying cross-sectional area is $A_d\sim 3 \times 10^{-3}$~cm$^2$ and the resistance is $R=\eta_d h_d /A_d \approx 0.01\,\Omega$. The energy released by Ohmic heating with a kA-level current over $t_p \sim 2 h / c \approx 3$~ps is therefore $\lesssim 0.1\,\mu$J. 

Will the conducting layer melt if it is supplied with $0.1\,\mu$J of heat energy? The volume of heated material is $h_d A_d \sim 10^{-2}$~cm$^3$, which gives a heated mass of $\sim 80$~mg. The available electrostatic energy is not sufficient to heat such a large mass, so it is unlikely that the melting of gas nozzles is caused by Ohmic heating from a discharge current.

\subsection{Heating caused by Plasma Deposition}

Plasma ions are the primary source of nozzle heating. Assuming that $N_i$ ions in the plasma are expanding isotropically, the angular ion flux is $N_i / 4\pi$.  The solid angle subtended by the nozzle is given by $2\pi(1-\cos\alpha)$, where $\tan\alpha=r_d/h$. For $h=0.5$~mm and $r_d=5$~mm, the solid angle is approximately $2\pi$ and the number of ions incident on the nozzle is around half the ion population.

The stopping power of 0.4~MeV protons in Cu is 166~MeV~cm$^2/$g\, \cite{PSTAR} and the rate of energy deposition is $\sim 1500$~MeV/cm. The ions therefore deposit all their energy over a distance $\lesssim 10~\mu$m. The maximum heated volume is $8\times10^{-6}$~cm$^3$ and the mass of heated material is $\sim 0.02$~mg. 

Based on the estimates of plasma density and volume from Sec. \ref{sec_Vulcan_params}, there are $N_i\sim 5\times 10^{14}$ ions at energy $\sim0.4$~MeV for a combined energy of 40~J. Half of these ions strike the nozzle and deposit an energy of 250~kJ/g. This is more than 100 times higher than the energy required to melt Cu, so plasma heating melts the nozzle easily. 

The same conclusion applies to a ceramic nozzle, even with the greater heat resistance. Consider SiC with a melting temperature of 3100~K, 370~J/g latent heat of melting and a specific heat capacity of $0.67-1.4$~J/gK for temperatures between $300-4000$~K \cite{NIST}. Using these values, we find that $\sim 4$~kJ/g is required to melt SiC. According to the NIST proton stopping catalogue, the stopping power of a 0.5~MeV proton in SiO$_2$ is 290~MeV~cm$^2$/g \cite{PSTAR}. For a density of 3.2~g~cm$^{-3}$, the rate of energy deposition is 1000~MeV/cm, and the deposition depth is $\sim 5\,\mu$m, for a heated volume of approximately $2 \times 10^{-5}$~cm$^3$ and a heated mass of $0.06-0.1$~mg. This is consistent with SRIM Monte Carlo simulations \cite{SRIM} of 0.4~MeV protons incident on SiC, which give a $4\,\mu$m range. 20~J of incident ion energy deposited in a few microns corresponds to heating of $\sim200$~kJ/g - more than sufficient to melt the ceramic. 

This conclusion is supported by a gas jet experiment on the VEGA-3 laser \cite{Henares_2023}. The laser was focused to an intensity of $2.2 \times 10^{19}$~W~cm$^{-2}$ in a He (97\%) and H$_2$ (3\%) gas mix with peak atomic densities of order $\sim 10^{21}$~cm$^{-3}$ (see Sec. \ref{sec:henares} for more details of the experiment). Tungsten nozzles suffered progressive damage over tens of shots when the laser was focused at a distance $\sim900\,\mu$m, whereas a UV fused silica nozzle was destroyed when the laser was focused at a distance of $400\,\mu$m from the nozzle surface. 

We apply this model of nozzle damage to a Vulcan-TAP experiment. The amount of energy deposited by the ions is sensitive to the number of ions and their average energy. Taking a vertical gas density profile from measurements by Hicks \textit{et al.} \cite{Hicks_2021}, we estimate the number of ions generated in the laser channel when the laser is focused at different heights above the gas jet. We assume a value of $N_i$ at a fixed height of $160\,\mu$m, where the density is maximal, and scale it by the normalised vertical density profile as the laser focus is shifted to different heights above the nozzle. Following the method outlined earlier, we then calculate the ion flux on the nozzle surface and the amount of energy deposited when the laser is focused at different distances from the nozzle. In the experiment, metallic nozzles were instantly destroyed at a nozzle-focus distance of $400\,\mu$m and virtually undamaged when they were separated by 2~mm. Since the melt threshold for Cu is 1.4~kJ/g, this suggests that the ion energy deposition must reach $\sim2$~kJ/g when the laser is focused somewhere between these two distances. These observations give bounds on the total number of 0.5~MeV plasma ions produced by the laser at a height of $160\,\mu$m, which satisfies $4\times10^{12} < N_{i} < 2\times10^{14}$. The product of the initial plasma volume and peak gas density gives a maximum electron number of $\sim 5 \times 10^{14}$, which may be considered an upper limit on the number of accelerated ions. 

\begin{figure}[!ht]
\centering
\includegraphics[width=0.4\textwidth]{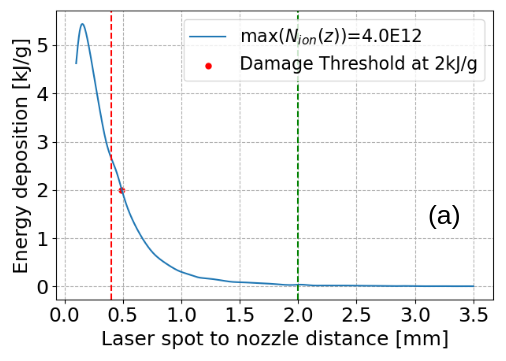} \\
\includegraphics[width=0.4\textwidth]{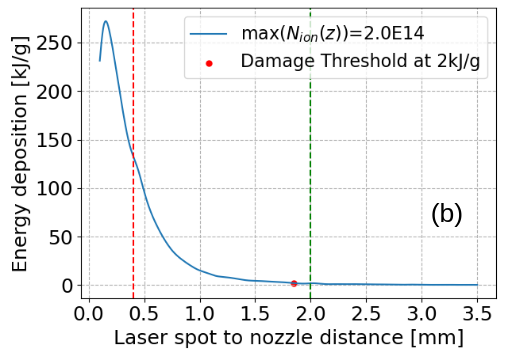}
\caption{Ion energy deposited per mass of nozzle material as a function of laser-nozzle separation for the number of ions $N_{i} = 4\times10^{12}$ (a) and $2\times10^{14}$ (b) normalised to the plasma height of $160\,\mu$m. The red dot represents the theoretical melt threshold for a Cu nozzle. Red and green vertical dashed lines represent observed distances where a steel nozzle was destroyed and survived, respectively. The ion energy is 0.5~MeV.}
\label{fig_nozzle_damage_threshold}
\end{figure}

The situation is illustrated graphically in Fig. \ref{fig_nozzle_damage_threshold}. Panel (a) shows the ion energy deposited in the nozzle as a function of distance, assuming $4\times10^{12}$ plasma ions at 0.5~MeV when the laser is focused at a height of $160\,\mu$m. Panel (b) shows the energy deposition for $2\times10^{14}$ plasma ions at a height of $160\,\mu$m above the nozzle. The amount of nozzle heating increases as the laser approaches the peak gas density at $\sim 160\,\mu$m and then drops steeply as the laser is focused further away from the nozzle. This is caused partially by the reduced solid angle subtended by the nozzle and also the changing gas density profile. Accounting for ion collisions in the gas before they reach the nozzle means that less overall energy is deposited in the nozzle, but the energy per unit mass of heated material increases due to a higher ion stopping power at lower energies. 

The estimates presented in this section suggest that nozzle damage is caused primarily by ion energy deposition rather than a resistive current, so nozzle material has little impact on nozzle survival. On the other hand, just as for EMP emitted from solid targets, the discharge current and EMP amplitude and spectrum change when one moves from conducting to dielectric nozzles. 

\section{Experimental Results} \label{sec_experiment}

\subsection{Ion Acceleration Experiment on VEGA-3}\label{sec_e1}

An experiment was conducted on the VEGA-3 laser system with the aim of using petawatt laser pulses to trigger laser channelling in inert gases (He, N$_2$, He-N$_2$) and thereby learn more about ion acceleration mechanisms \cite{Ospina_2024, Ospina_2022}.

\begin{figure}[!ht]
\centering
\includegraphics[width=0.4\textwidth]{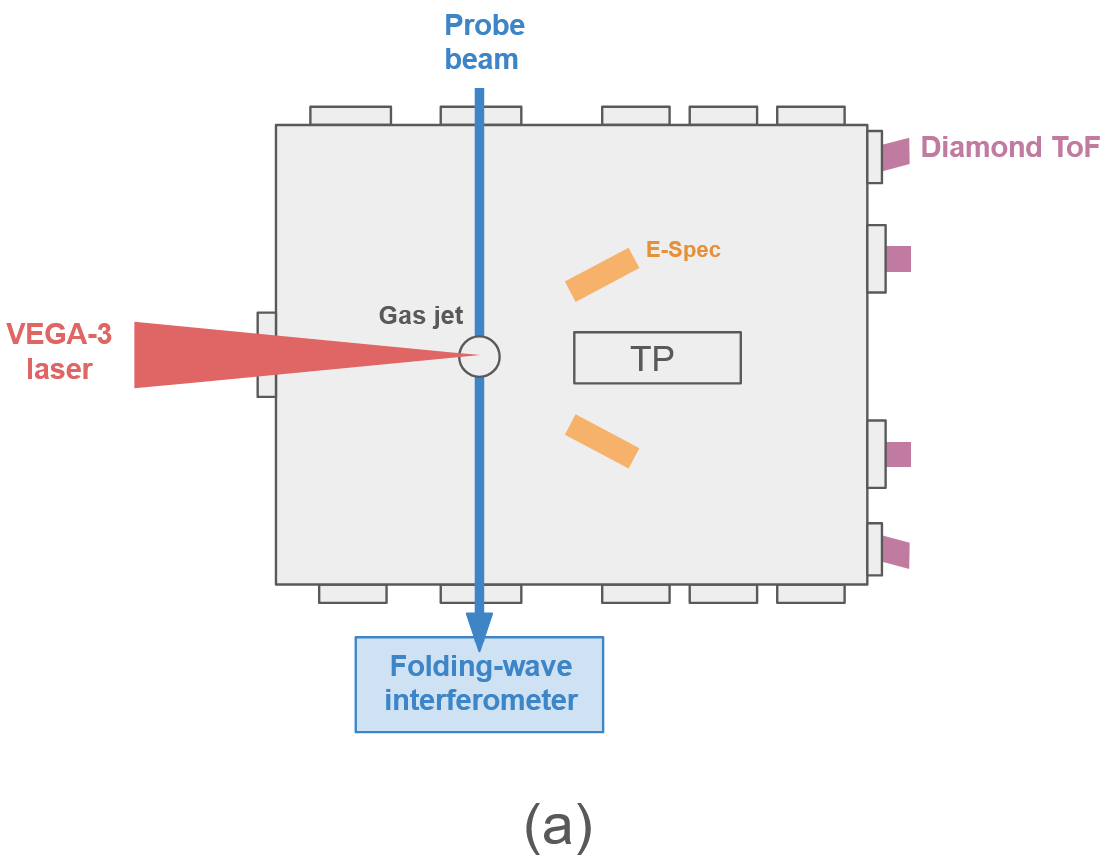} \\
\includegraphics[width=0.4\textwidth]{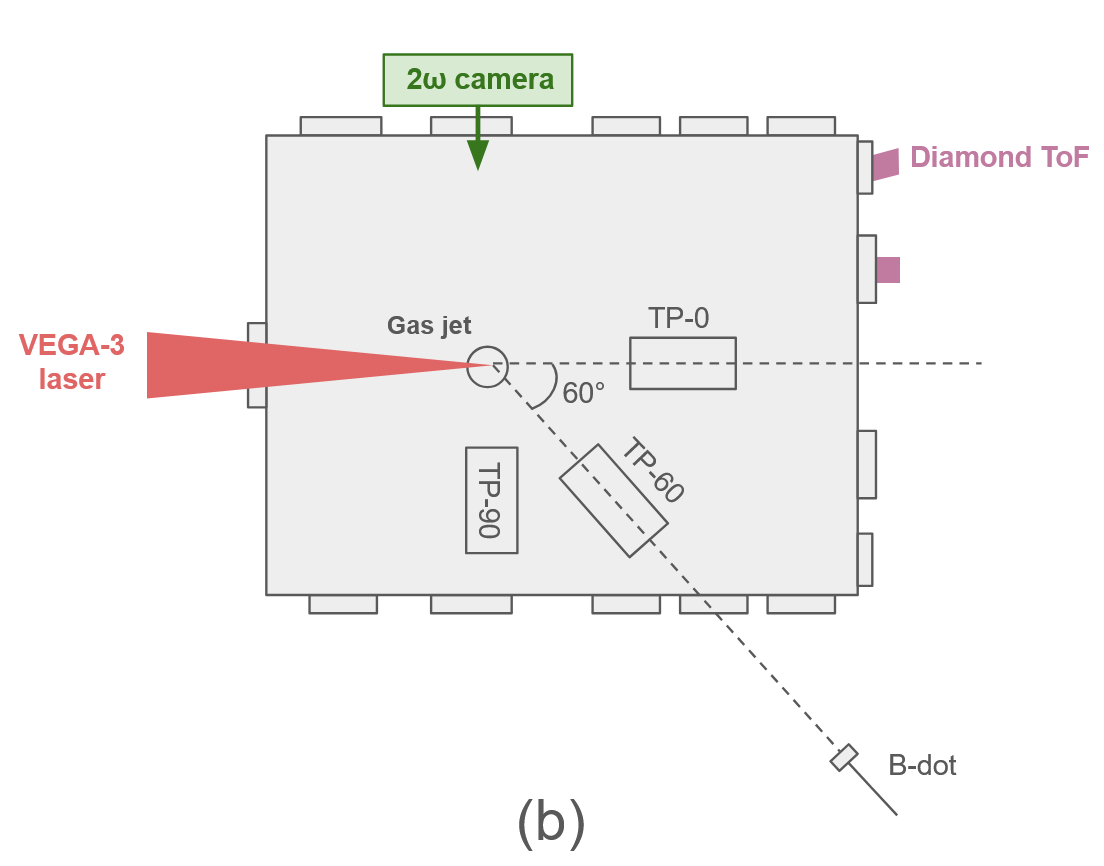}
\caption{Schematic diagrams of two experiments conducted on the VEGA-3 laser system. (a) Set-up described in Sec. \ref{sec_e1}. A Thomson parabola spectrometer (TP), two electron spectrometers (E-Spec) and four diamond time-of-flight detectors (Diamond ToF) are represented by coloured boxes. (b) Set-up described in Sec. \ref{sec:henares}. Three Thomson parabola spectrometers are placed at $0^{\circ}$, $60^{\circ}$ and $90^{\circ}$ to the laser axis. The B-dot probe was positioned at variable distances ($\sim 2-3$~m) from the gas target, with its measurement axis horizontal and orthogonal to the line-of-sight axis. A camera was used to take images of the gas at twice the laser fundamental frequency. Further details of these experiments can be found in Refs. \cite{Ospina_2024, Henares_2023}.}
\label{fig_VEGA3_expts}
\end{figure}

A two-lens imaging system was used to monitor the focal spot diameter during the experiment, which was fixed at $\sim 12~\mu$m FWHM. A laser pulse duration of $\tau_L=72\pm~24$~fs was measured using an autocorrelator. The laser energy varied in the $18.4\pm 2.3\,\rm J$ range (uncertainties correspond to the standard deviation over $\sim 100$ shots), though only $\sim$47\% of the pulse energy is contained within the focal spot\cite{Ospina_2022}. Two types of shock nozzle were used: Sourcelab J2021 nozzles (peak atomic density $n_{\rm at,max} \approx 5 \times 10^{20}$~cm$^{-3}$ at a height of $\sim 500~\mu$m above the nozzle) and S900 nozzles \cite{Henares_2019} (peak atomic density $n_{\rm at,max} \approx 10^{20}$~cm$^{-3}$ up to $\sim 900$~$\mu$m above the nozzle). The S900 nozzles were designed to produce a convergent shock further from the nozzle surface than the J2021 design in order to reduce nozzle damage and degradation of the gas density profile. Interferometric measurements showed that the J2021 had a narrower, denser gas profile along the laser axis than the S900 nozzle. They also revealed that He gas produced denser wings and broader density peaks than N$_2$ or the He-N$_2$ mix. J2021 and S900 nozzles will hereafter be referred to as ``short-" and ``long-focus" nozzles, respectively.

\begin{figure}[ht]
\centering
\includegraphics[width=0.4\textwidth]{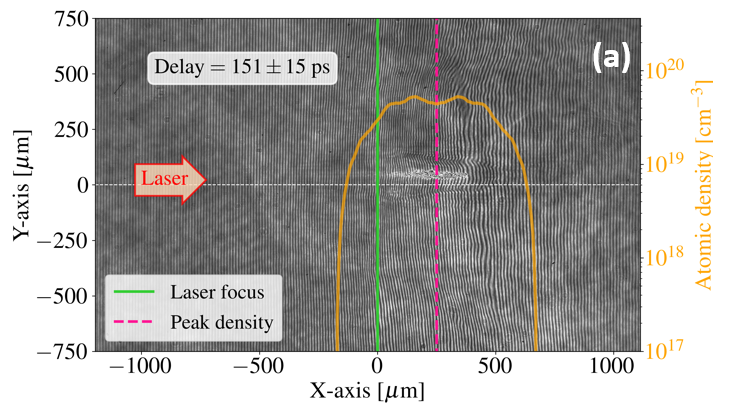} \\
\includegraphics[width=0.4\textwidth]{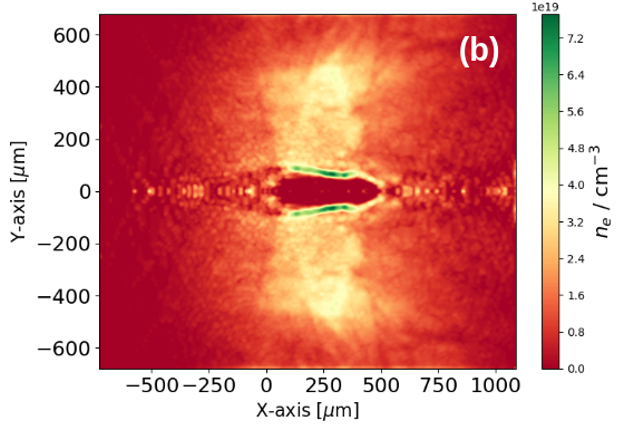}
\caption{Images corresponding to optical probe arrival $\sim 150$~ps after the drive laser in a He gas with a long-focus shock nozzle (a) Raw interferogram (b) Density map showing a plasma channel straddling the peak density region at $x = 250$~$\mu$m. The laser focus position in the vacuum was set at $x=0$~$\mu$m.}
\label{s147}
\end{figure}

To measure the electron density in the plasma channel, a folding-wave interferometer was used to take optical interferograms of the ionised gas at different temporal delays relative to the drive laser \cite{Pisarczyk_2019}. Figure \ref{s147}b shows a He density distribution from the experiment taken $\simeq 150$~ps after the arrival of the laser, revealing a laser channel with average radius $\simeq 70~\mu$m. Interferograms could not be recovered at earlier probing times because the channel boundaries were obscured by plasma self-emission. Modelling the expansion as a Sedov-Taylor cylindrical blast wave \cite{Hutchens_1995, Zeldovich_2002}, $R(t)=\xi_0(\mathcal{E}_bt^2/\rho_{\rm gas}l_p)^{1/4}$, gives an energy of $\mathcal{E}_b\simeq 1.3$~J (taking $\xi_0=0.9$ \cite{Zeldovich_2002}) deposited in the channel over its length of $l_p\simeq 0.5$~mm, which is less than 20\% the deposited laser energy of 7.3~J.  Such a significant difference is explained by the fact that the expanding plasma is essentially collisionless. Plasma ions propagate through the ambient gas and ionise atoms without creating a density compression - only a small portion of the slow ion population contributes to shock wave formation.  

2D PIC simulations in CALDER \cite{Ospina_2024, Debayle_2017} and experimental interferograms of the gas at the point of laser arrival suggest that, by the time the laser has travelled through the low-density wings of the gas density profile and arrived in the central high-density region, the laser has become significantly defocused. For a gas density of 4.5~bar and a plasma channel radius of $20~\mu$m, the plasma volume is $\sim6\times 10^{-7}$~cm$^3$, the laser energy deposited in the gas is 7.3~J, the bulk electron temperature is $T_e \simeq 140$~keV and the pressure is $\sim50$~Mbar, see Tab. \ref{tab1c}). 

The stopping power of 140~keV electrons in He is $\sim3$~MeV~cm$^2$/g\,\cite{ESTAR}, which corresponds to a mean free path of more than 50~cm for our gas density. The Debye radius, however, is of order $0.1~\mu$m, so bulk electrons remain in the plasma due to the Coulomb attraction and the plasma expansion is quasi-neutral. Bremsstrahlung losses are negligible on this timescale, so the plasma expands adiabatically.

Assuming that all bulk electron energy is transferred to ions, the average energy of helium ions is about 0.43~MeV, their velocity is $\sim 5\,\mu$m/ps and the expansion time is 100~ps. In this experiment, about 10\% of the absorbed laser energy is transferred to hot electrons with an effective temperature of $\sim$1.1~MeV given by the ponderomotive scaling. Then using the plasma charging model described in Sec. \ref{sec_model}, we estimate a plasma capacitance of 0.0086~pF, a plasma potential of 4.4~MV and 42~nC charge of escaped electrons. The plasma electrostatic energy is about 0.1~J and the escaped electrons have carried away $\sim0.15$~J.

Applying the analysis of gas ionisation described in Sec. \ref{sec_model}, we find that the photoionisation probability is very low - about $1.6\times 10^{-7}$ - the level of ionisation by fast electrons is two orders of magnitude larger at $\sim 3.3\times 10^{-5}$ and the dominant ionisation processes are electrical breakdown and ions in the expanding plasma. The plasma ionisation time is, therefore, within a few tens of ps. The plasma discharge results in EMP emission with energy $\sim 0.17$~mJ, which corresponds to an electric field of $\sim 6.3$~kV/m and a magnetic field of $21\,\mu$T at 1~m from the source. Nozzle damage is due to the energy deposition of a few joules by the plasma ions. 

\begin{figure}[!ht]
\centering
\includegraphics[width=0.4\textwidth]{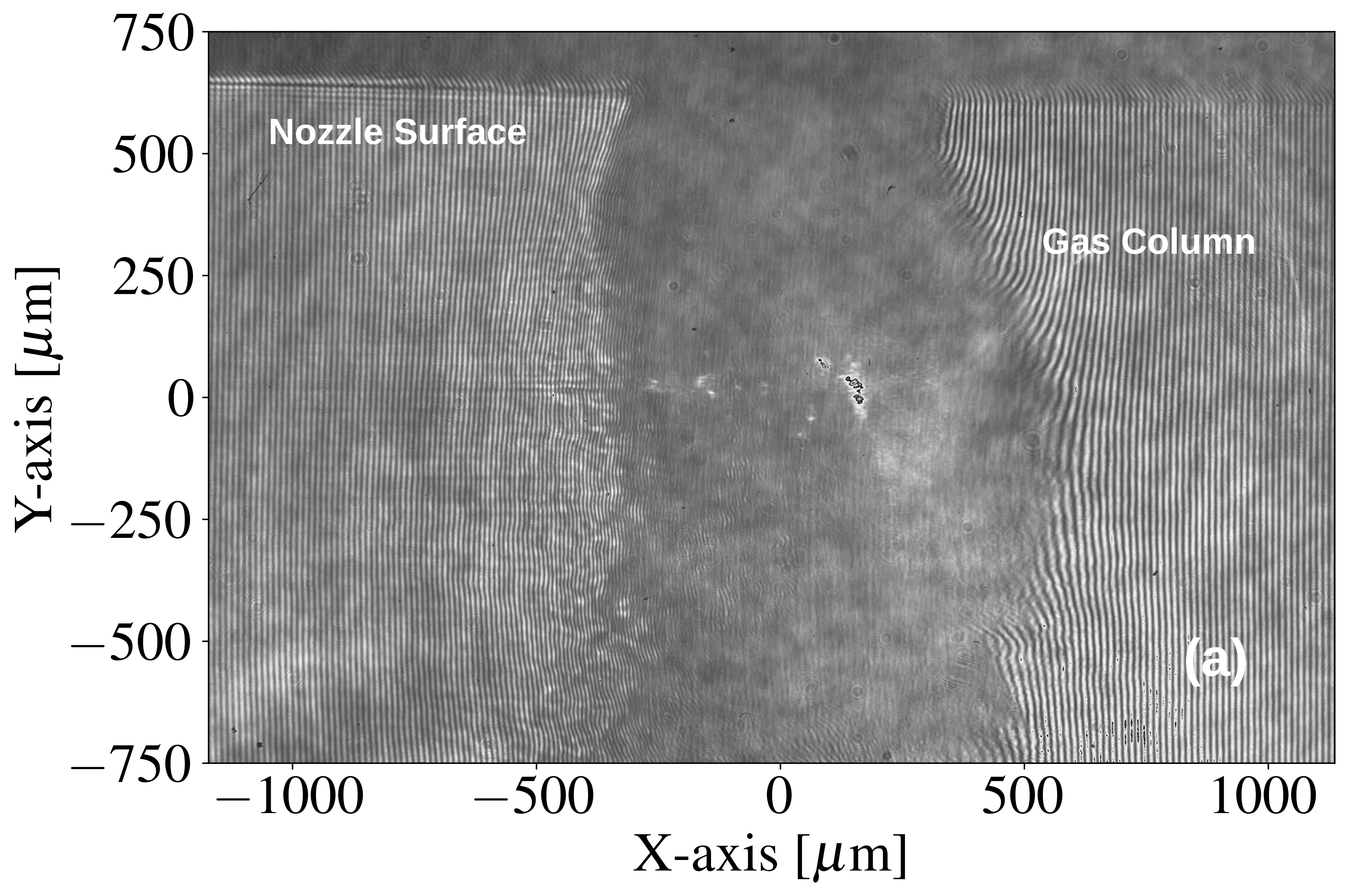} \\
\includegraphics[width=0.4\textwidth]{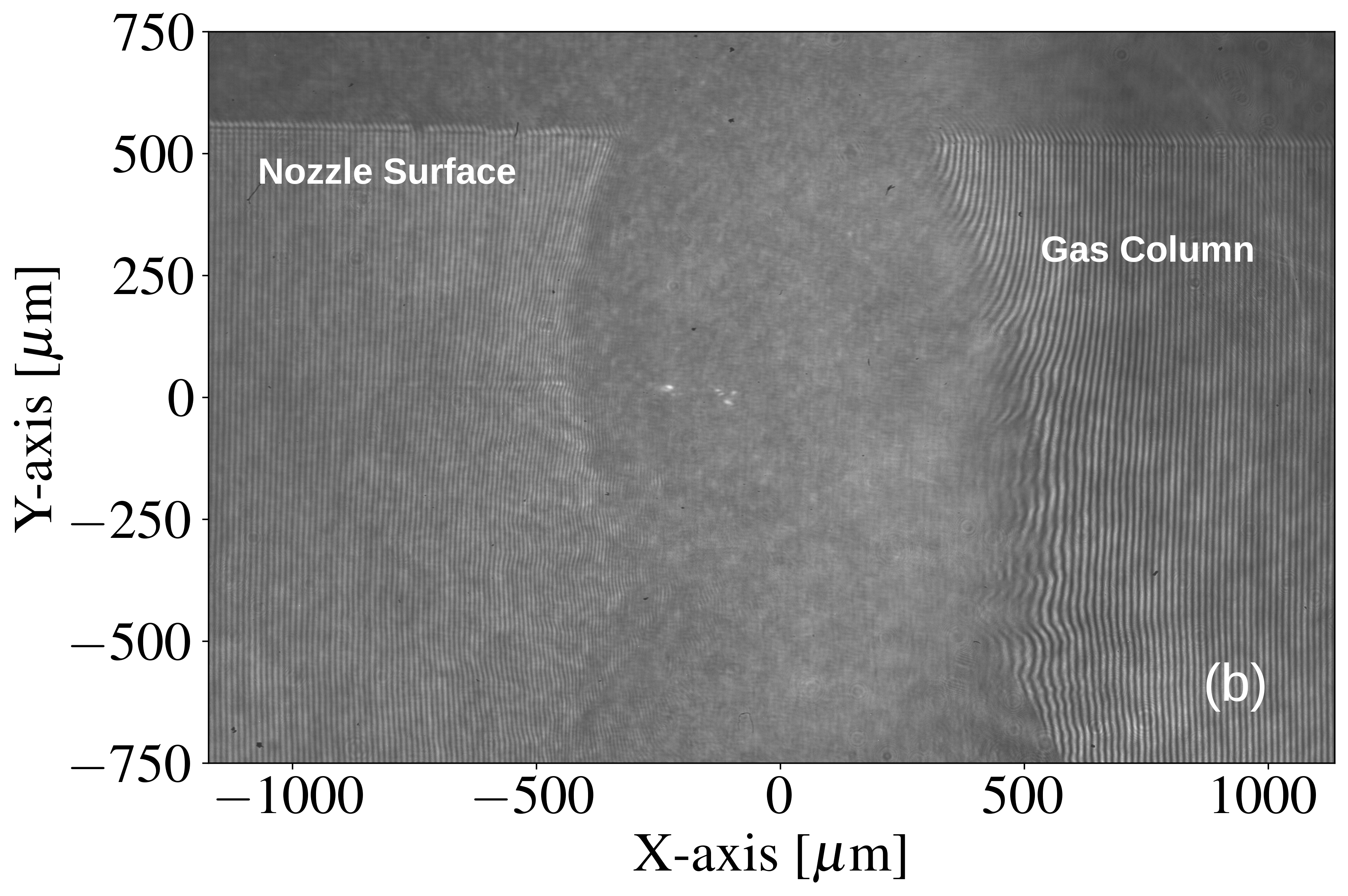}
\caption{Interferograms of the VEGA-3 laser interacting with an N$_2$ gas ejected from a short-focus shock nozzle. Probe times relative to the arrival of the pump beam are 40~ps (a) and 90~ps (b). The laser intensity is $10^{20}$ W/cm$^2$ and the gas density is $10^{20}$ cm$^{-3}$.}
\label{fig_N2_time_scan}
\end{figure}

Progress of gas ionisation between the laser and nozzle was measured with a pick-off beam converted in the second harmonic and directed transversely to the laser channel for interferometric imaging. Figure \ref{fig_N2_time_scan} shows interferograms taken at two different times after the arrival of the laser pulse. Panel (a) corresponds to the earliest probing time measured after the laser pulse arrival. The shadow of the gas is visible in both images, suggesting that the entire gas volume can be ionised within 40~ps. The speed of the ionisation front, therefore, exceeds $\sim 20\,\mu$m/ps, which is consistent with gas ionisation by electrical breakdown.

To monitor ion emission during the experiment, five diamond time-of-flight detectors \cite{Salvadori_2020} were placed in multiple different locations around the vacuum chamber ($0^\circ$, $\pm 17^\circ$, $4.85^\circ$, $5.5^\circ$ with respect to the laser axis, always inclined at $9^\circ$ to the horizontal plane of the laser) and a Thomson parabola spectrometer was placed along the laser axis in the same horizontal plane as the laser. When the diamond detectors were placed at $\pm 17^\circ$, nitrogen ions with energy up to $\sim 80$~MeV were measured. For shots on He gas, $\alpha$-particles were accelerated to 0.7~MeV/u with $10^9$ particles per steradian, while shots on N$_2$ gas produced ions up to 5.7~MeV/u with $10^7$ particles per steradian. These ions were measured in the laser forward direction within a $\pm 17^\circ$ cone \cite{Ospina_2024}. Sample He spectra from the time-of-flight diagnostics can be found in Fig. \ref{fig_ion_spectra}b. We expect higher yields perpendicular to the laser axis.

\subsection{EMP Experiment on VEGA-3} \label{sec:henares}

A second experiment was conducted on VEGA-3 with a view to characterising the EMP and ions produced in laser-gas interactions. The laser energy before compressor was $\sim 30$~J and the pulse duration was 30~fs. Based on images of the focal spot taken at low energy, 21\% of the energy on-target was contained within the first Airy disk \cite{Henares_2023} for an on-target intensity of order $\sim 10^{20}$~W~cm$^{-2}$ at best compression, similar to the intensity reported for the experiment in the previous section. The laser contrast was $10^{-12}$ at $0.1$~ns, with no significant pre-pulses \cite{Henares_2023}. A Prodyn RB-230(R) radiation-hardened B-dot probe was used to measure the amplitude of the magnetic field at different distances from the gas jet, while two diamond time of flight detectors and three Thomson parabolas were used to monitor charged particle emission. Signal degradation due to EMP pick-up in the unshielded oscilloscope prevented the easy identification of a photo-peak in the time of flight data, so quantitative analysis of the ion spectrum is not possible. A proton spectrum from the Thomson parabola positioned at $90^{\circ}$ to the laser axis is shown in Fig. \ref{fig_ion_spectra}c. The data is integrated over five shots, with a lower proton sensitivity limit of $\sim 1.3$~MeV. The proton yield increases towards lower energies, with a maximum of $\sim 3\times10^{11}$~MeV$^{-1}$sr$^{-1}$ at the low-energy limit. B-dot signals were recorded on a Rohde\&Schwarz RTO64 digital oscilloscope with 2~GHz bandwidth and 10~GS/s sample rate. The B-dot probe was oriented so that its axis was parallel to the ground and sensitive to an azimuthal magnetic field relative to the nozzle axis. 

B-dot voltage signals were bandpass-filtered between 0.4~GHz and 2~GHz, with the lower limit determined by the probe frequency response \cite{Raczka_2017, Dubois_2018} and the upper limit by the oscilloscope bandwidth. The waveforms were then cropped, the zero-point offset removed, and cable attenuation corrected in the Fourier domain. The cable attenuation functions were measured using a Rohde\&Schwarz ZNA4 Vector Network analyser from 5~MHz to 4~GHz, but the correction was only applied between 0.4~GHz and 2~GHz to avoid amplifying noise beyond the sensitivity range of the diagnostic. Finally, the voltages were integrated in time and multiplied by the probe effective area of 0.2~cm$^2$ (supplied by the manufacturer) to yield the magnetic field.

\begin{figure}[ht]
\centering
\includegraphics[scale=0.4]{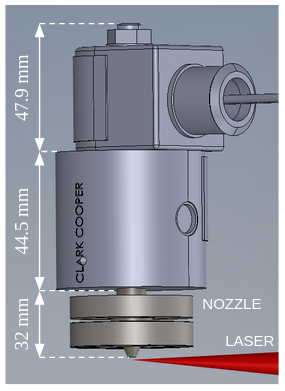}
\caption{3D graphic of the gas jet nozzle and solenoid valve assembly at VEGA-3. Arrows indicate dimensions relevant to electromagnetic emission.}
\label{fig_VEGA3_target}
\end{figure}

\begin{figure}[ht]
\centering
\includegraphics[scale=0.4]{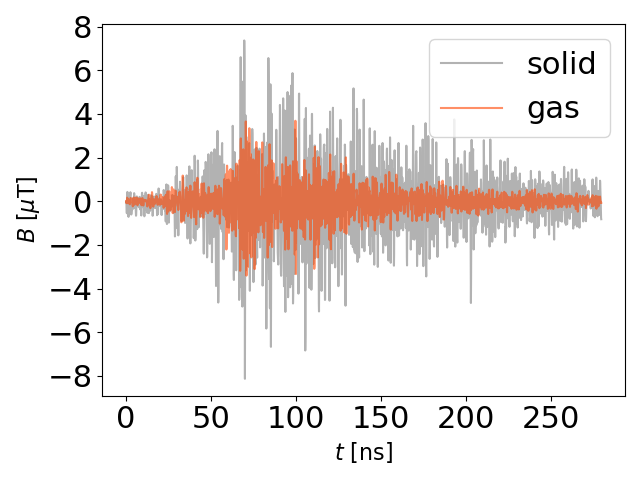}
\caption{Comparison of EMP waveforms for solid and gaseous targets on VEGA-3. The signals were measured using a Prodyn RB-230(R) probe positioned at $60^{\circ}$ to the laser forward direction at a horizontal distance of $2.66$~m from the nozzle and vertically in-line with the laser focal spot. The maximum amplitude of the magnetic field was a factor $2-3$ times lower for the gas targets compared to $6~\mu$m-thick solid Al foils.}
\label{solid_vs_gas}
\end{figure}

Both solid and gaseous targets were studied during the experiment. The gas was a 3:97 H$_2$-He mix at $\sim 1000$~bar backing pressure, forced vertically downwards through the nozzle aperture to the laser focus. The electron density in the laser focal region was close to $10^{21}$~cm$^{-3}$. On shots with solid targets, the nozzle and solenoid valve assembly (Fig. \ref{fig_VEGA3_target}) was replaced by a bracket and multi-foil array. The foils were made from $6~\mu$m-thick aluminium. Figure~\ref{solid_vs_gas} shows typical EMP waveforms for solid and gaseous targets, demonstrating a reduction by a factor $2-3$ in the peak magnetic field for the same laser parameters when solid targets were switched to gas. This is consistent with the results of Kugland \textit{et al.} \cite{Kugland_APL_2012} on the PHELIX laser with an Ar gas jet and tallies with our theoretical estimates. The ChoCoLaT2 code \cite{Poye_2018} evaluates the charge produced by the VEGA-3 laser interacting with $6~\mu$m-thick Al foils to be of order 100~nC, which is twice higher than the 42~nC estimated for a gas in Sec. \ref{sec_e1}.  

\begin{figure}[ht]
\centering
\includegraphics[width=0.4\textwidth]{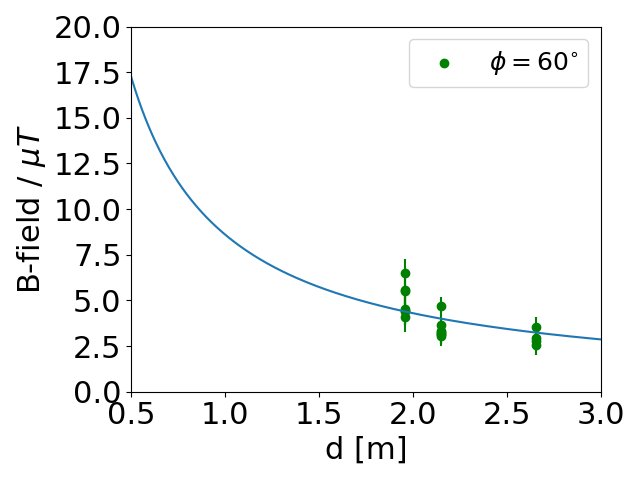}
\caption{Variation of EMP maximum magnetic field with distance from the gas jet. Data was collected with the B-dot probe positioned at $\phi = 60^{\circ}$ to the laser axis, with the line of sight to the target occluded. The fitted curve is for a 3~cm-tall antenna with the angle between the antenna axis and the probe assumed constant at $\theta = 90^\circ$ for the different distances.}
\label{distance_scan}
\end{figure}

\subsubsection{Comparison with EMP Emission Model}

Figure~\ref{distance_scan} shows measurements of the decay of the magnetic field with distance from the gas jet. The data points come from experimental measurements and represent the average of the maximum and minimum magnetic field values in the waveform. Error bars represent one standard deviation from the mean. The green points represent data taken when the B-dot was positioned at $60^{\circ}$ to the laser axis, outside the vacuum chamber and behind a glass window. A Thomson parabola spectrometer inside the chamber blocked the probe's direct line of sight to the target. 

The magnitude of the magnetic field is of order $5-10\,\mu$T at a distance of $\sim 2$~m from the jet. Significant shot-to-shot variations are due to adjustments in the laser focal position inside the gas to optimise the laser coupling to the target. There are insufficient data points to identify an unambiguous scaling with distance, but they can allow us to estimate the charge accumulated in the gas by the laser ejection of hot electrons if we assume a dipolar radiation field. The solid line represents a least squares fit to the data points with Eq. (1) from Ref. \cite{Minenna_2020}, assuming an antenna height of $h_d=3$~cm (consistent with the vertical height of the jet nozzle in Fig. \ref{fig_VEGA3_target}) and leaving the target charge $Q$ as a free parameter. The best-fit charge is $\sim 20$~nC in this case. This is in agreement with the model presented in Sec. \ref{sec_model}, which predicts a charge of $\sim 40$~nC for the VEGA-3 experimental conditions. 

Peaks in the EMP Fourier spectrum constrain the EMP emission mechanism. The average of the EMP Fourier spectra for shots with the gas jet targets reveals multiple prominent resonances between $\nu_{\rm emp} \sim 1.2$~GHz and 1.9~GHz (not shown here). Assuming antenna emission from a monopole of height $c/4\nu_{\rm emp}$, the strongest resonances correspond to heights of $\sim 4-5$~cm - similar to the dimensions of the nozzle. Note that our measurements were only sensitive to emissions from 0.4~GHz to 2~GHz and there were other large metallic objects in the chamber that might contribute to the EMP.       

\section{Discussion} \label{sec_discussion}

\begin{figure}[!ht]
\centering
\includegraphics[width=0.35\textwidth]{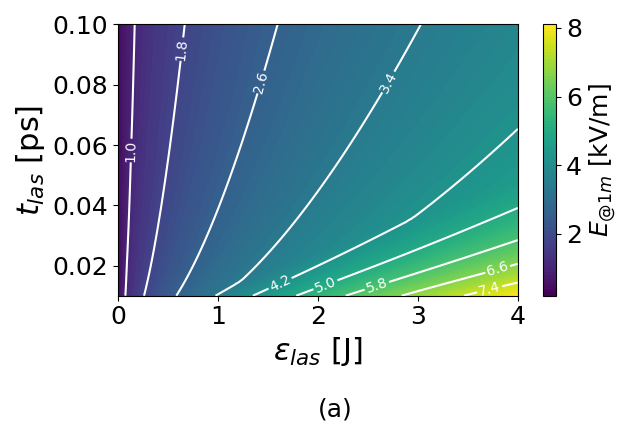} \\
\includegraphics[width=0.35\textwidth]{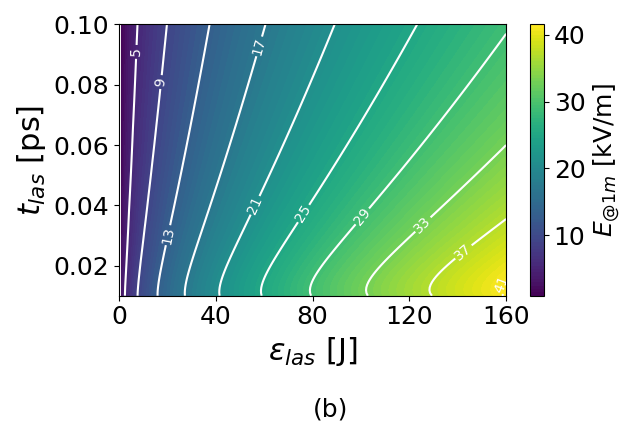} \\
\includegraphics[width=0.35\textwidth]{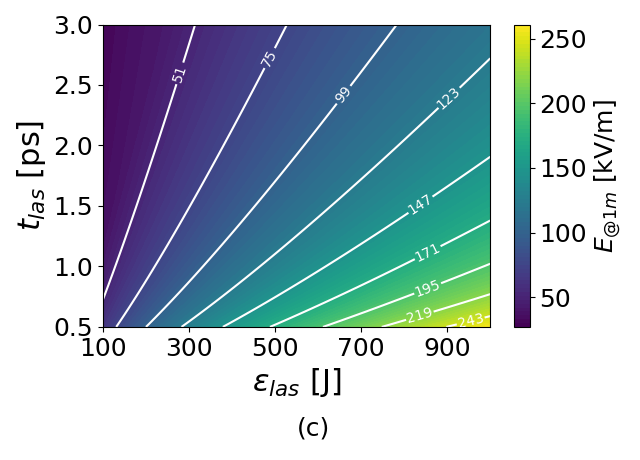}
\caption{Variation of the EMP electric field located 1~m from a 3~cm-tall nozzle for different values of the total laser energy and pulse duration, calculated using the model from Sec. \ref{sec_model}. Laser and gas parameters have been chosen so that they are representative of experiments with: (a) under-dense gases on low-energy systems like the Gemini Target Area 2 laser \cite{Scott_2021}, (b) near-critical gases and PW-scale lasers, such as VEGA-3, (c) under-dense gases and high energy, longer pulse duration lasers such as LMJ PETAL\cite{Batani_2013} and NIF ARC\cite{DiNicola_2015}.}
\label{fig_EMP_scan}
\end{figure}

We examine here how EMP amplitude and nozzle damage vary on different laser systems. In Fig. \ref{fig_EMP_scan}, we consider the EMP produced by three different types of laser-gas interaction: (a) experiments at relatively low laser energy with under-dense gases, (b) experiments with PW-scale lasers and near-critical density gases, (c) experiments with high energy, longer pulse duration lasers and under-dense gas targets.

For all the calculations shown in Fig. \ref{fig_EMP_scan}, we assume 40\% of the total laser energy shown on the $x$-axis is contained within the focal spot and that all the laser energy in the focus is absorbed over the full length of the plasma channel. The gas is He, ejected from a 3~cm-tall conducting nozzle, and the laser-to-hot-electron conversion efficiency is assumed to be 20\% in panel (a), 30\% in panel (b) and 40\% in panel (c). For the relatively low-energy laser systems in panel (a), we assume a plasma channel radius of $r_p = 20\,\mu$m, channel length $l_p = 3$~mm and a gas pressure of $p_g = 0.2$~bar, with the laser focused $h_p = 3$~mm above the nozzle. In panel (b), we take $r_p = 20\,\mu$m, $l_p = 0.5$~mm, $p_g = 10$~bar and $h_p = 0.5$~mm. Then for the PETAL-type lasers in panel (c), we consider $r_p = 40\,\mu$m, $l_p = 1$~cm, $p_g = 0.2$~bar and $h_p = 3$~mm. 

Calculations suggest that low-energy lasers can eject a few tens of nC from a plasma channel and produce EMP fields of a few kV/m at 1~m from the target. PW-class lasers are more disruptive, producing tens to hundreds of nC plasma charge and EMP fields of several tens of kV/m at 1~m from the gas. The strongest fields are expected for PETAL-class lasers, which can displace hundreds of nC to $\mu$C of charge and produce EMP fields of several hundred kV/m at metre-scale distances. These estimates agree well with recent measurements by W. Cayzac \textit{et al.} at PETAL \cite{Cayzac_2024}, which confirm that the EMP fields are generally a factor few lower than you would expect for solid target interactions. 

The magnitude of the EMP can be controlled either by disrupting the discharge current as it propagates down the gas nozzle or through its relation to the amount of charge that escapes the plasma. A higher laser energy increases the energy available for conversion into hot electrons, while a higher intensity will increase the hot electron temperature. Increasing the laser energy and intensity will therefore tend to increase the escaping electron charge. A larger plasma volume or higher gas density will spread the laser energy over more particles and generally lower the average particle energy. A larger volume will also increase the plasma capacitance and reduce the electrostatic barrier potential for the escaping electrons.

\begin{figure}[!ht]
\centering
\includegraphics[width=0.4\textwidth]{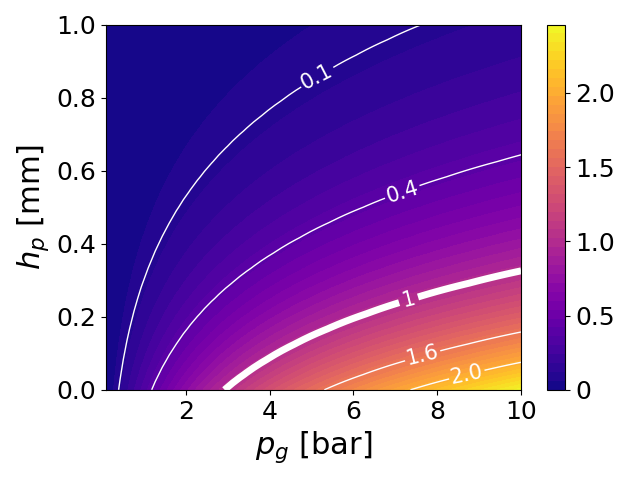}
\caption{Nozzle damage factor (ratio of the plasma ion energy deposited in the nozzle per mass of heated material divided by the nozzle melt threshold) for different values of the laser-nozzle distance and gas pressure, calculated using the model from Sec. \ref{sec_model}. An exponentially decaying gas density profile with $500$~$\mu$m scale length is assumed, where the peak He gas density is taken at the nozzle surface ($h_p=0$). A bold white line marks where the nozzle damage factor is equal to 1, corresponding to a deposited energy-per-mass equal to the melt threshold of Cu ($\sim 1.3$~kJ/g).}
\label{fig_nozzle_damage_scan}
\end{figure}

Our model can also be used to estimate when a nozzle is likely to be damaged on a given laser system. By way of example, Fig. \ref{fig_nozzle_damage_scan} shows the damage produced by a VEGA-3 type laser focused at different heights, $h_p$, above a metallic nozzle. The gas pressure, $p_g$, is varied up to $10$~bar and the different colours correspond to the ion energy deposited in the nozzle as a fraction of the energy required to melt the nozzle ($\sim 1.3$~kJ/g). Any contour value above $1.0$ therefore implies that the nozzle will be damaged. An artificial, exponentially decreasing density profile with $500\,\mu$m scale length is assumed, while the other parameters are as given in Table \ref{tab1c} for the VEGA-3 experiment. Damage occurs for laser-nozzle distances below a few hundred microns and peak gas pressures above a few bar. Damage is most easily avoided by focusing the laser further from the nozzle, where the gas density is lower and the ion flux on the nozzle will be reduced. 

It is important to bear in mind some of the limitations of our model. Many of the physical quantities used as inputs depend on each other in complicated ways, such as the relationship between the plasma channel volume and laser focal spot size, or between gas pressure and the hot electron conversion efficiency. Without additional information about these connections, the model will not be able to accurately predict how the EMP fields scale with physical parameters. Our model also does not account for plasma acceleration processes like shock \cite{Haberberger_2012} or laser wakefield \cite{Mangles_2004} acceleration, which can affect the electron and ion distributions in complicated ways.

\section{Conclusions} \label{sec_conclusion}

We have explored various mechanisms of gas ionisation and nozzle damage in laser-gas interactions. Depending on the gas density and plasma temperature, an electrical discharge and EMP can be triggered either by plasma expansion from the laser focus or electrical breakdown of the ambient gas. The relatively small plasma volume produced in laser-gas interactions leads to a small plasma capacitance and this in turn produces a smaller ejected charge and lower level of EMP compared to solid targets. Nozzle damage is caused by plasma ions depositing energy in the nozzle surface rather than Ohmic heating from the discharge current. Two experiments on the VEGA-3 laser have provided useful supporting evidence for our model. The minimum speed of gas ionisation - inferred from optical interferometry - is consistent with collisional ionisation by ions in an expanding plasma or with the electric discharge.

Further dedicated experiments are needed to confirm the model presented in this paper, in particular by measuring the emitted ion and electron spectra over a large solid angle. Measuring the average ion energy will allow us to estimate the discharge time and plasma expansion velocity. It is also important to measure the laser-ejected charge since 10 times more hot electrons imply 10 times stronger EMP fields. Direct measurement of a discharge current in the nozzle and characterisation of the EMP spectrum is essential to confirm that the EMP is related to antenna emission from the nozzle.

EMP sources couple with the chamber and objects within it, exciting resonant modes \cite{Mead_2004}. These modes are excited differently according to the nature of the driver and may last much longer than the source itself, depending on their relaxation time. We have discussed a plasma discharge as one important source of EMP, but the electron beam ejected from the plasma can also induce currents and secondary electromagnetic fields in the nozzle and surrounding chamber \cite{Consoli_2021}. The intensity of the induced current is much smaller than that of the escaped electrons, however, and its impact on the nozzle damage is expected to be small. 

Experimental validation of the nozzle damage mechanism requires measurement of the ion energy distribution in the radial direction, in the energy range below 1~MeV. This might also be combined with a study of nozzle damage as the laser is focused at various different heights above the gas nozzle. Switching from metallic to ceramic nozzles would be useful for assessing the impact of nozzle material.  


\begin{acknowledgements}
The authors thank the staff at the VEGA-3 Laser Facility for their hard work over the course of several experiments. This work has been carried out within the framework of the EUROfusion Consortium, funded by the European Union via the Euratom Research and Training Programme (Grant Agreement No. 101052200 — EUROfusion). Views and opinions expressed are, however, those of the authors only and do not necessarily reflect those of the European Union or the European Commission. Neither the European Union nor the European Commission can be held responsible for them. We acknowledge GENCI for providing us access to the Joliot-Curie supercomputer (grants 2021-A0130512993 and 2022-A0130512993). We also acknowledge Grant PID2021-125389OA-I00, funded by MCIN/ AEI / 10.13039/501100011033 / FEDER, UE and ``ERDF A way of making Europe'', funded by the European Union. V. Ospina-Boh\'{o}rquez and C. Vlachos acknowledge support from the LIGHT S\&T Graduate Program (PIA3 Investment for the Future Program, ANR-17-EURE-0027). This work received funding from the European Union’s Horizon 2020 research and innovation program through the European IMPULSE project under grant agreement No 871161 and from LASERLAB-EUROPE V under grant agreement No 871124. It is published as part of the international project ``PMW”, co-financed by the Polish Ministry of Science and Higher Education within the framework of the scientific financial resources for 2021-2022 under the contract no 5205/CELIA/2021/0 (project CNRS no 239915). Finally, we acknowledge the financial support of the IdEx University of Bordeaux / Grand Research Program ``GPR LIGHT".
\end{acknowledgements}

\bibliography{references}

\end{document}